\documentclass[draft]{agujournal2019}
\usepackage{url} 
\usepackage{lineno}
\usepackage[inline]{trackchanges} 
\usepackage{soul}

\usepackage{amsmath} 
\draftfalse

\journalname{Geochemistry, Geophysics, Geosystems}

\begin{document}

%
%


\title{Morphology, interactions, and evolution of laminar thermal plumes revealed by time-resolved volumetric velocity measurements and analysis: Application to Earth's mantle}

%
%




\authors{Xiyuan Bao\affil{1}\thanks{Now at Department of Earth and Planetary Sciences, Harvard University, Cambridge, MA, USA}
Carolina Lithgow-Bertelloni\affil{1}}

\affiliation{1}{Department of Earth, Planetary, and Space Sciences, University of California, Los Angeles, Los Angeles, CA, USA}





\correspondingauthor{Xiyuan Bao}{xiyuanbao@g.ucla.edu}



\begin{keypoints}
\item Time-resolved 4D velocity measurements reveal collective dynamics of laminar thermal plumes in high-Prandtl-number convection.

\item Plumes show behaviors including merging, splitting, branching, pulsing, and detachment, from inter-plume interactions and background flow.

\item Establish a network of plume evolutionary pathways, a benchmark for geodynamical models, help link surface to source processes.
\end{keypoints}

%
%

%
%


\begin{abstract}
Plume-laden convection plays a fundamental role in thermal transport in geophysical and industrial flows. While the dynamics of isolated laminar plumes have been extensively studied, their collective interactions and evolution in a multi-plume system remains poorly quantified due to the lack of four-dimensional (4D, in space and time) flow measurements for viscous convection, common in the turbulence community. Here we present a time-resolved experimental study of clustered plume dynamics in a high-Prandtl-number Rayleigh-B\'enard system with temperature-dependent viscosity from 4D velocity measurements. Building on our companion methodology \cite{bao2025piv}, we characterize the spatiotemporal distribution, morphological evolution, and interaction modes of tens of coexisting plumes. We identify a wide spectrum of dynamical behaviors---merging, splitting, branching, pulsing, and head detachment---that arise from coupled interactions among plumes, multiscale flow, and the evolving thermal boundary layers. We synthesize prior behaviors into a unified, time-resolved framework of plume evolutionary pathways. Characteristic length and timescales, including spacing and initiation times, are compared with previous theoretical and experimental studies. We find thinner plumes with greater minimum spacing. The fraction of isolated plumes (e.g., Hawaii-like) varies dramatically (0–60\%) as the flow evolves. Our results provide new insight into plume self-organization and spatiotemporal variability under Earth-like conditions, crucial for understanding geochemical evolution and mapping geochemical anomalies from hotspot lavas to deep sources. These results quantitatively complement previous experimental work and numerical studies on interacting plumes, serving as a benchmark for comparison with geodynamical simulations of mantle convection.
\end{abstract}

\section*{Plain Language Summary}
Hot, buoyant plumes play a major role in how Earth’s mantle moves heat and material, but their behavior is typically studied one at a time. In reality, multiple plumes often coexist and interact. In this study, we use precise laboratory experiments and advanced flow analysis to track how dozens of plumes evolve over time in a viscous fluid heated from below---mimicking mantle conditions. We find that plumes frequently merge, split, branch, pulse, and detach their heads, forming a dynamic network of interacting structures. By quantifying plume spacing, size, and motion through time, we show that textbook ``isolated plumes" are rare and short-lived. Our results reveal a diverse range of plume behaviors and evolutionary pathways shaped by boundary layer instabilities, background flow, and interaction between plumes. This experimental work provides the first comprehensive laboratory, time-resolved map of plume dynamics in a convecting high viscosity fluid offering a new reference for interpreting numerical models and understanding the sources and surface expressions of plumes.

%
%
\section{Introduction}
Laminar plumes, or focused, buoyant upwellings that efficiently release heat and material to the surface, are expected to be present under vigorous convection such as that found on Earth's and other planetary mantles. Thermal plumes anchored at the core-mantle boundary have been long suggested to be responsible for volcanic hotspots like Hawaii \cite{Morgan1971} and other ocean islands. The origin, dynamics, and material transfer in plumes remains of great interest to the Earth science community, because of their connection to structure imaged in the deep interior \cite{French2015} and as probes into its composition and geochemical evolution. The motion and kinematics of plumes and hotspots may also offer first-order constraints on absolute plate motions \cite{Morgan1972,Jellinek2004}. 

A clear pathway to understanding how to map geochemical anomalies from ocean island basalts to their source in Earth's interior remains elusive, despite prior extensive and pioneering work \cite <cf.> {Davaille2015treatise}. There have been some fundamental technical challenges to further progress. A complete surface-to-source mapping requires knowledge not only of single plume origin, morphology, and material entrainment, but also of the interactions of the plume with the surrounding flow. Both experimentally and computationally, this has been difficult. Experimentally, because visualization techniques did not provide enough quantitative measurements of the 4D flow, nor were the analysis tools at a stage where both individual plume structure, plume evolutionary pathways, and ambient flow dynamics could be separately quantified. Computationally, even with the advances of the last decade, it has been challenging to achieve the resolution required to resolve small-scale and large-scale flow and their interactions at the same time \cite{mohr2023challenges}, let alone material tracking that can resolve Lagrangian Coherent Structures \cite<LCSs, >{Shadden2011,Haller2015}. Hence, much prior work, discussed below, has by necessity focused on individual plumes, or large-scale flow. More limited, but still insightful, have been studies of plume interactions with each other or the surrounding fluid or plume morphological evolution through their life cycle \cite<e.g.,>{olson1985creeping,Griffiths1990,Gonnermann2004,Davaille2005,Cagney2015}. 

Extensive efforts have been made towards understanding the dynamics of mantle plumes in the laboratory, where plumes have been treated from spherical diapirs \cite{Manga1993,Kelly1997}, Rayleigh-Taylor instabilities \cite{whitehead1975dynamics}, injected lighter fluid \cite{olson1985creeping,Griffiths1990}, to instabilities from spot/line heating \cite{shlien1976some,forstrom1967experiments,Kaminski2003,Cagney2015} or basal heating (Rayleigh-B\'enard convection, \citeA{Sparrow1970,lithgow2001plume,Davaille2005}). These studies focused on finding scaling relationships for different stages in the plume cycle, from initiation to full development, spreading and even collapse \cite{Androvandi2011,Cagney2016b}. Others focused on interactions between laminar thermal plumes \cite{Moses1993,Kelly1997,schaeffer2001interaction} and plume clustering \cite{Gonnermann2004}. Various other complexities have been considered experimentally, including thermochemical plumes and convection in stratified fluids \cite{Davaille2002,Jellinek2002,lebars2004whole,kumagai2008mantle}, spherical geometry \cite{futterer2012isoviscous}, non-Newtonian rheology \cite{Davaille2013}, visco-elasto-plastic fluids with subduction \cite{davaille2017experimental}, and internal heating \cite{Limare2019}.Many aspects of plume dynamics have been also explored in numerical simulations, where other physical phenomena relevant to Earth's mantle are better studied, such as the effects of phase changes, compressibility and depth-dependent properties, and plate-like boundaries  \cite{Leng2012,Arnould2020}.

Despite this substantial body of work, there are no published quantitative 4D velocity fields (3D + time) for multiple interacting laminar plume at conditions applicable to planetary interiors (high Rayleigh number $Ra$ and Prandtl number $Pr$), partly due to complications arising from optical distortion \cite{bao2024self} and the need for a more complex velocimetry system and postprocessing. Unlike in the turbulence community where 4D fluid flow can be routinely measured \cite{westerweel2013particle,schroder20233d}, including scanning-based methods \cite{hori2004high,casey2013scanning,tapiasilva2024high} applied to turbulent Rayleigh-B\'enard convection \cite{fujisawa2005scanning,fujita2020three}), to the best of our knowledge, this is the first attempt at such measurements in viscous fluids. Previous visualization of these laminar plumes are typically available within a projected view and not in the perpendicular direction \cite{Griffiths1990,lithgow2001plume,Davaille2005}. Based on the system in \citeA{Newsome2011}, \citeA{Cagney2015} performed experiments and measured 4D velocity and temperature, but their study was limited to a single axisymmetric thermal plume.

Numerical simulations offer flexible control and tracking of Earth-like convective flow, however, simultaneously resolving both small-scale instabilities like plumes and large-scale flow in 4D remains challenging \cite{Lin2006a,kronbichler2012high,mohr2023challenges}. The quantitative comparison between laboratory and numerical results, necessary to ground truth numerical work, has been also largely limited to a single axisymmetric thermal plume \cite{Vatteville2009,Cagney2016a}. 

The lack of 4D flow measurements in a simple, viscous Rayleigh-B\'enard convection prohibited holistic, quantitative investigation of the spatiotemporal distribution and evolution of plumes, and the interactions amongst plumes and with the ambient fluid. The extraction of static and dynamic information at the individual plume level from the multi-scale flow is also impossible without a tailored, dedicated, end-to-end methodology \cite{bao2025piv} to identify and track individual plumes and plume clusters objectively. As a result, most previous studies have examined only isolated aspects of plume-plume interactions or the life cycle of a single plume, whereas much less attention has been paid to the collective evolution pathways that arise in the presence of clusters of plumes \citeA{Davaille2015treatise}. A summary of previous work can be found in Table 1 in \citeA{bao2025piv}.

Here, we build upon prior work spanning the theoretical foundations of high-Rayleigh-number boundary-layer convection \cite{Batchelor1954,Howard1966,Sparrow1970,whitehead1970thermal,whitehead1975dynamics}, the dynamics and scaling of isolated thermal plumes \cite{Moses1993,Manga1999,Davaille2005,Davaille2011,Newsome2011,Cagney2015}, the influence of compositional layering and rheology on plume behavior \cite{Jellinek2002,LeBars2004,Solomatov2006,Androvandi2011}, and numerical investigations of plume morphology and plate-coupled mantle flow \cite{parmentier2000three,labrosse2002hotspots,Leng2012,Arnould2020}, and report results based on our experimental 4D velocimetry measurements of plume clusters, using 3D Scanning, Stereoscopic, Particle Image Velocimetry (SSPIV) and the methodology described in \citeA{bao2025piv} to analyze the flow using LCSs and cluster analysis. We are thus able to outline a network of evolutionary pathways for plumes for future applications to Earth’s mantle, derived from \textit{quantitatively} examining the full flow in space and time, extending the scope of previous experimental and numerical studies.  

Our goal and hope is that the results provided here, enabled by the end-to-end methodology in \citeA{bao2025piv} provide a cornerstone for future surface-to-source mapping of geochemical anomalies in hotspot lavas and for the interpretation of the more complex structure illuminated by recent seismic imaging \cite<e.g.,>{French2015,Tsekhmistrenko2021,dongmo2023imaging}.

We have chosen the simplest possible system with good dynamical similarity to the Earth's mantle, i.e., a basally heated Newtonian fluid with temperature dependent viscosity in a rectangular tank, with Earth-like convective vigor where inertial forces are negligible ($Ra=1.9 \times 10^6, Pr>3000$; local $Ra>10^8$ defined below, and Ra=$10^6$ is required to be in the plume regime, \citeA{Davaille2005}). Like much of the previous laboratory and numerical work we build upon, our experiment is formulated as an initial-value problem: plume formation and evolution are studied as a controlled transient response in a fixed rectangular geometry, rather than as a complete analogue of Earth's four-billion-year convective history. As noted above, it does not include the plate-like boundaries, and therefore lacks the associated plate-driven lateral flow of the mantle and of plumes, and shear beneath tectonic plates. Its purpose is to isolate and quantify plume-scale morphology, interaction, entrainment, and evolutionary pathways under viscously dominated, high-Prandtl-number conditions, providing a benchmark for comparison with more complete geodynamic models. The experimental design and tailored analysis methodology necessary to fully process the data and extract measurements relevant to plume structure, and interactions can be found in \citeA{bao2024self,bao2025piv}. In this study, we will just briefly summarize the experimental setup and analysis tools (section \ref{sec:setup}) needed for interpretation. We first describe the distribution pattern of the plumes in the 4D space and the associated timescales and length scales (section \ref{sec:distribution}). We then identify the morphological evolution and rich dynamical behaviors of clusters of plumes, enabled by our 4D velocity data and Lagrangian analysis (section \ref{sec:plume_behaviors}). We focus on the network of plume evolutionary pathways, which emerges from combining all distributional and dynamical information, and impacts how surface hotspots may be mapped to their source locations. Finally, we compare our results with previous studies, and discuss implications for the Earth.

\section{Experimental setup and analysis methodology}
\label{sec:setup}

The experiment was carried out by heating tracer-laden corn syrup in a plexiglass tank. The internal dimensions of the tank are 405 mm $\times$ 275 mm $\times$ 405 mm along $x,y,z$ ($y$ is positive downward). The bottom of the tank was uniformly heated at $t=0$ s. The temperature, $T$, increased essentially linearly from room temperature 25 $^\circ$C to 80 $^\circ$C in 250 s and remained constant until the end of the experiment ($\sim8000$ s). The lid of the tank was kept at $T=25^\circ$C. Two cameras moved simultaneously with an LED Light source along $z$ . The LED light source illuminated a thin (5 mm) light sheet in the $x,y$ plane. The cameras were used to extract both in-plane and out-of-plane displacement from particle images (cf. Fig. 1 in \citeA{bao2025piv}).

SSPIV data was acquired for 69 tank planes at 5 mm interval along the $z$ axis (Fig. \ref{fig:velocity_ftle_cluster}). We acquired a total of 60 epochs (Epoch 0 to 59) of data for the duration of our experiment. Each epoch, a complete velocity scan of the entire tank, had a duration of $\sim$130 s, with two frames scanned for each plane. A group of two non-overlapping neighboring planes were scanned and immediately rescanned, with an inter-frame interval of $<1.5$ s. The final 4D velocity data at different planes have been synchronized via cubic interpolation and are available from Epoch 0 to 58, covering $68\%$ of the fluid volume ($>86\%$, $>92\%$ and $>85\%$, along $x,y,z$ respectively), with $3.53 \times 3.53$ mm in-plane (extendable to 0.2 mm, \citeA{bao2024self}) and 5 mm inter-plane resolution.

The viscosity $\eta$ of our working fluid is temperature-dependent:
$\eta = \mathrm{exp}(4.642 \times 10^{-4} T^2 - 1.246 \times 10^{-1} T + 6.325)$ ($T$ in $^\circ$C, $\eta$ in Pa$\cdot$s), so that the viscosity contrast is as much as 65 under our experimental conditions. The global Rayleigh number is $Ra = 1.9 \times 10^6$ for the highest viscosity, and nearly two orders of magnitude larger for the lowest viscosity at the base, i.e., $Ra_\delta = 1.2 \times 10^8$, comparable to Earth's mantle (Table \ref{tab:properties}). The Prandtl number varies from $Pr=2.2 \times 10^5$ at 25 $^\circ$C to $Pr=3.4 \times 10^3$ at 80 $^\circ$C. While the $Pr$ for the mantle is essentially infinite ($10^{22}$), our Prandtl number is still large enough to safely neglect the effects of inertia \cite{Whitehead2013}. As an additional check, the maximum Reynolds number is $<<$ 1 ($Re=0.78$), using the measured characteristic upwelling velocity of 1 mm/s (cf. section \ref{sec:distribution}) and minimum viscosity at the base $\eta(T=80^\circ$C). Hence, the experimental results described below are dominated by viscous forces and negligibly affected by inertia. This experimental condition ($Ra > 10^6$, $Pr>>10^3$) spans the regime transition where boundary-layer thickness, plume spacing and $Ra$ scalings \cite{Androvandi2011} can be tested directly. The temperature-dependent viscosity further allows us to examine how such relations evolve as viscosity varies through more than one order of magnitude.

Using a diffusive timescale $\tau_\mathrm{diff}=H^2/\kappa$ ($7.2\times 10^5$ s in the experiments and $9\times 10^{18}$ s for the mantle), where $H$ is the thickness of the fluid, and $\kappa$ the thermal diffusivity, Table \ref{tab:properties}), we can scale the duration of the experiment to the mantle. Every second corresponds to 0.4 Myr of mantle convection, and the entire experiment ($\sim8000$s) corresponds to $\sim$ 3200 Myr, representative of mantle evolution. Alternative timescale based on the rising speed, as short as 0.037 Myr for each second in our experiment, will be discussed in Sec. \ref{sec:discuss_speed}.

\begin{table}
\caption{Comparison of convective properties between mantle and laboratory conditions$^a$.}
\label{tab:properties}
\begin{tabular}{@{\extracolsep\fill}lccr}
\hline \hline
Properties  & Symbol & Mantle & Laboratory \\[3pt]
\hline
Temperature jump & $\Delta T$ &  3000 K & 55 K\\
Thickness & $H$ & 3000 km & 0.275 m \\
Density & $\rho$ & 4000 $\mathrm{kg/m}^3$ & 1427 $\mathrm{kg/m}^3$ \\
Dynamic viscosity & $\eta$ & $10^{20} \sim 10^{23}$ Pa$\cdot$s$^b$ & $0.51\sim 33.1$ Pa$\cdot$s$^c$ \\
Thermal expansion coefficient & $\alpha$ & $3\times 10^{-5}$ K$^{-1}$ & $4.1\times 10^{-4}$ K$^{-1}$ \\
Thermal diffusivity & $\kappa$ & $1\times 10^{-6}$ m$^2$/s & $1.05\times 10^{-7}$ m$^2$/s \\
Prandtl number & $Pr = \frac{\eta/\rho}{\kappa}$ & $10^{22}\sim2\times10^{25}$ & $3\times10^3\sim2\times 10^5$ \\
Rayleigh number & $Ra = \frac{\rho g \alpha \Delta T H^3}{\eta \kappa}$ & $10^6\sim10^9$ & $2\times10^6\sim10^8$ \\
\hline \hline
\end{tabular}

$^a$By default, mantle properties from \citeA{Turcotte2002}; lab data for corn syrup at 25$^\circ$C .

$^b$Range from \citeA{Mitrovica2004}.

$^c$Lowest viscosity at 80$^\circ$C.
\end{table}

\begin{figure*}
  \centerline{\includegraphics[width=\linewidth]{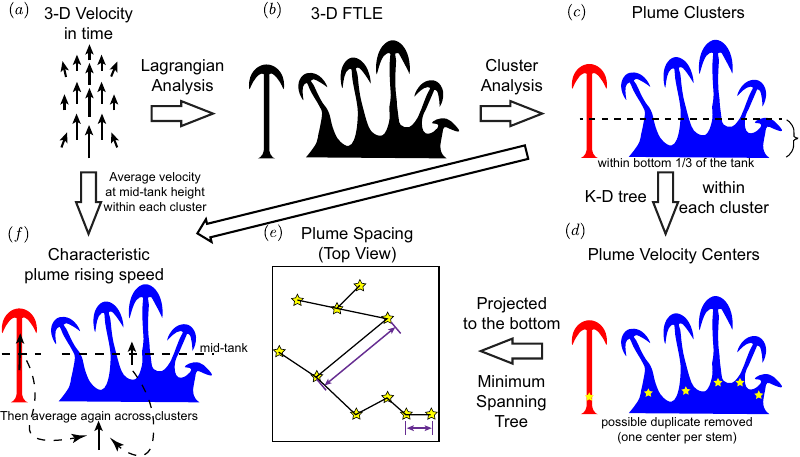}}
  \caption{Schematic of the SSPIV-based analysis workflow. The 4D velocity field in (a) is used to generate the backward Finite-Time Lyapunov Exponent (FTLE) field $\sigma_f$ in (b), via Lagrangian analysis, then FTLE ridges in the upwelling region are separated into clusters in (c). Within each cluster, plume centers, or upwelling velocity maxima as stars in (d) are identified in the bottom 1/3 of the tank (i.e., 91.7 mm, or the entire cluster if its total height is less than this threshold), using a K-D tree projected onto the tank bottom in (e), where plume spacing is determined by the minimum spanning tree, that connects all the plume centers using nearest neighbors, and the total length of the path is the shortest. (f) The average velocity at mid-tank height within each cluster is then averaged over all clusters to define the characteristic rising speed of the plume in each snapshot.}
\label{fig:analysis_schematic}
\end{figure*}

\begin{figure}
  \centerline{\includegraphics[width=\linewidth]{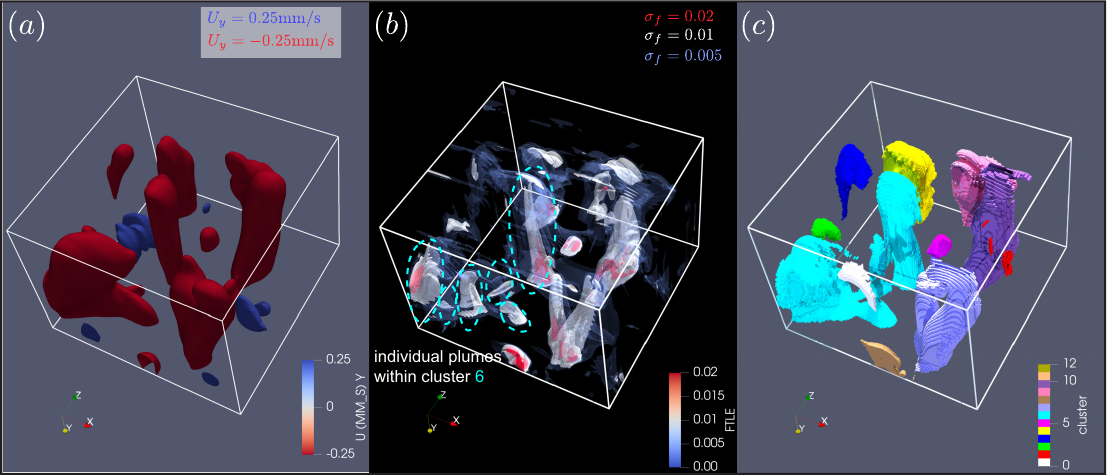}}
  \caption{3D visualizations of the plumes in Epoch 40. (\textit{a}) Vertical velocity $U_y$ contours of $\pm0.25$ mm/s. Note that the connectivity of the upwelling and downwelling depends on the contours visualized, cf. Movie S1 for further visualizations. (\textit{b}) Backward FTLE field $\sigma_f$ contours of 0.02, 0.01, 0.005 $\mathrm{s}^{-1}$. Individual plumes contained in the same cluster 6 are highlighted with cyan colored dashed line ovals. (\textit{c}) Plume clusters, each cluster contains either a single plume, or multiple interconnected plumes as indicated in Figure \ref{fig:analysis_schematic}c.}
\label{fig:velocity_ftle_cluster}
\end{figure}

The methodology specially designed to analyze the 4D velocity data \cite{bao2025piv} is summarized in Fig. \ref{fig:analysis_schematic}. With this methodology the 4D velocity can be recast into the Lagrangian frame as a backward Finite-Time Lyapunov Exponent (FTLE) field $\sigma_f$. The ridges corresponding to the plume material or high $\sigma_f$ regions with sharp edges, are isolated from the fluid domain and separated into disconnected plume clusters, and plume centers, from which we obtain plume spacing and characteristic plume rising speeds ($U_y$). We have included the visualization of the 4D vertical velocity in Movie S1, and the FTLE and plume clusters in Movie S2. A representative snapshot as discussed in \citeA{bao2025piv} at Epoch 40 is shown in Fig. \ref{fig:velocity_ftle_cluster} with all three fields.

As the FTLE ridges tend to be broader than in reality due to their limited temporal resolution ($\sim130$ s per epoch) \cite{bao2025piv}, we use them to determine plume locations and the morphology of plume clusters, for which they provide more robust information. For morphology and structure of individual plumes, such as stem head and width, where the temporal resolution is limiting, we use quantitative constraints from the tracers in the raw SSPIV images and from plume-induced optical distortion phenomena \cite{bao2025piv}. These image-based and optical-distortion measurements provide complementary and representative measurements on plume stem and head dimensions where suitable tracer or plume-shadow features are visible.

In addition to the methodology described in \citeA{bao2025piv}, we estimate the overall plume rising speed per epoch (denoted as $U_\mathrm{plume}$, the characteristic plume rising speed). We take the average mid-tank $-U_y$ ($y$ pointing down) in each plume cluster (over the cross-section of the cluster), and average them again over all clusters (Fig. \ref{fig:analysis_schematic}a, c, f). We exclude the first few epochs since plumes are still slow and accelerating, and they may not have yet reached the mid-tank height. After we obtain $U_\mathrm{plume}$, the corresponding plume travel time $\tau_\mathrm{plume}$ can be defined as $L_\mathrm{between}/U_\mathrm{plume}$, where $L_\mathrm{between}$ is the distance between the top and bottom thermal boundary layers (TBLs).



\section{Results: Spatiotemporal pattern of plumes}
\label{sec:distribution}

The measurements just described and further explained in \citeA{bao2025piv} allow us to provide direct measurements of the spatiotemporal distribution of plumes, their life cycles, dynamical behavior and evolutionary pathways. Our results can then be directly compared with theoretical scaling expectations for plume spacing and boundary layer thickness from previous studies \cite{Howard1966,Davaille2005}. Our analysis focuses primarily on hot upwelling plumes, while cold downwellings are shown and discussed only insofar as they affect hot-plume interactions.
\subsection{Distribution of plumes}
\label{sec:plume_distribution}
\begin{figure*}
  \centerline{\includegraphics[width=\linewidth]{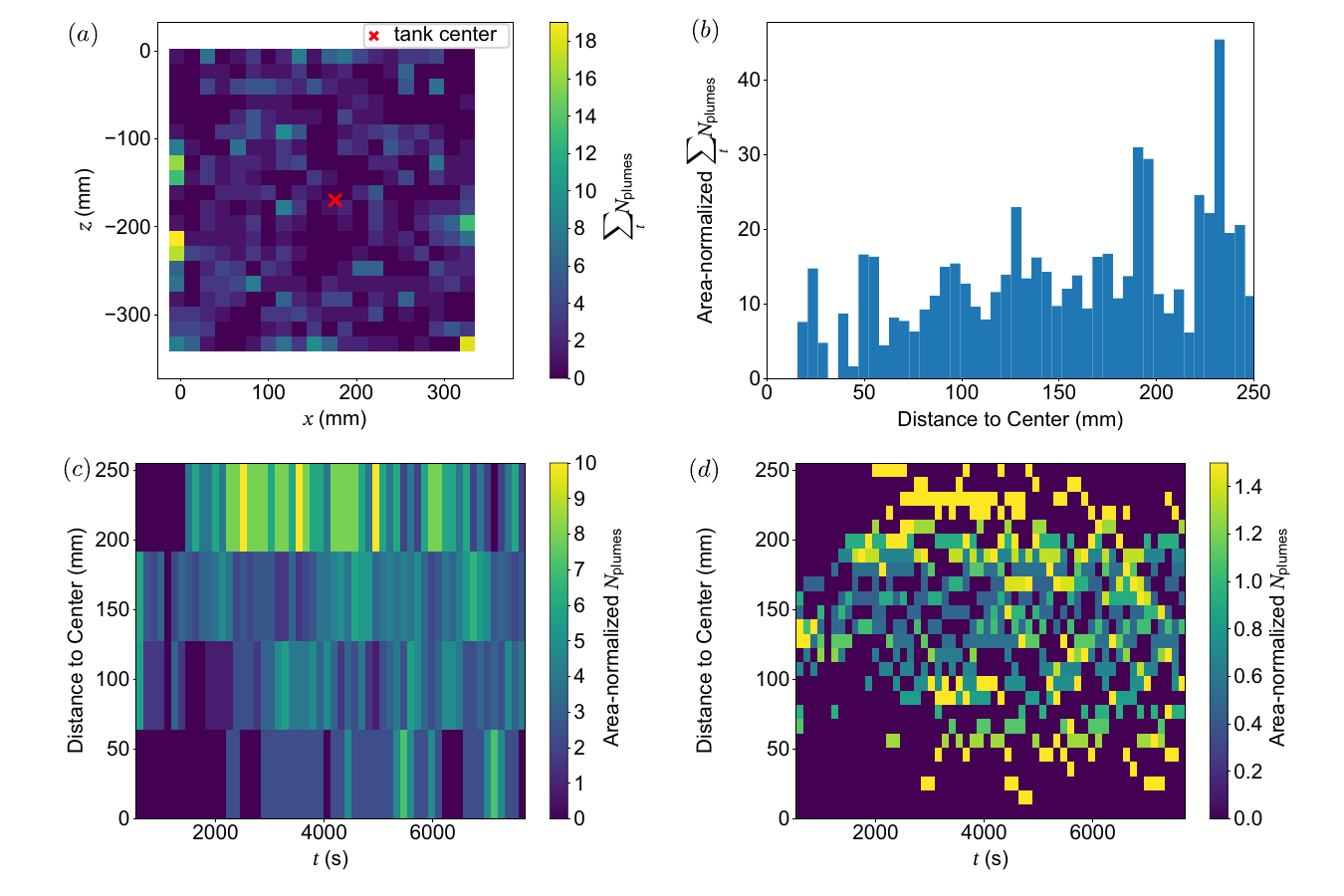}}
  \caption[The distribution of plumes]{The distribution of plumes in space and time. (\textit{a}) 2-D histogram of $N_\mathrm{plume}$ along $x$-$z$, the lateral plane, and integrated over all time. The red cross shows the center of the tank. The boundary of the plot indicates the side walls of the tank. $y$ is positive pointing into the plane. (\textit{b}) 1-D histogram of $N_\mathrm{plume}$ integrated over all time against the distance to the center of the tank (i.e., each bin is a 2-D annulus). Numbers of plumes are normalized based on the observed plume occupied area within each distance bin. Area-normalized spatial-temporal $N_\mathrm{plume}$ distribution is shown with different chunks in (\textit{c}) and (\textit{d}).}
\label{fig:hist}
\end{figure*}

We stack the 1-D (distance from the tank center in $x-y$) and 2-D ($x-y$) histograms of plume centers across all epochs covering the duration of our experiment (Fig. \ref{fig:hist}a, b). We do not observe clear spatial preferences for plume formation, except that the total number of plumes ($N_\mathrm{plume}$) is higher near the walls (Fig. \ref{fig:hist}a, b). Interestingly, no plume originates within 20 mm of the tank center.

The temporal distribution of plume numbers as a function of distance from the tank center is shown in Fig. \ref{fig:hist}c, d. In Fig. \ref{fig:hist}c the distance from the center is binned every $\sim$64 mm and (Fig. \ref{fig:hist}d, every $\sim10$ mm. Existing or new plume centers migrate back and forth amongst different distance bands (green to yellow blocks in Fig. \ref{fig:hist}c, d), at various timescales (from 100s to 1000s seconds, cf. section \ref{sec:timescales}). We find plumes initiate in a narrow distance range from the tank center (125$\pm$10 mm), then gradually spread, to nearly all distances from the center after $\sim3000$ s. While $N_\mathrm{plume}$ tends to be higher as we approach the tank walls for most of the experiment duration, it decays to average values towards the end of the experiment. More plumes form near the center (within $\sim60$ mm distance) over time . This increase is clear in Fig. \ref{fig:hist}c after $\sim5000$ s. 

\subsection{Number of plumes}
\label{sec:numbers}
\begin{figure}
  \centerline{\includegraphics[width=1\linewidth]{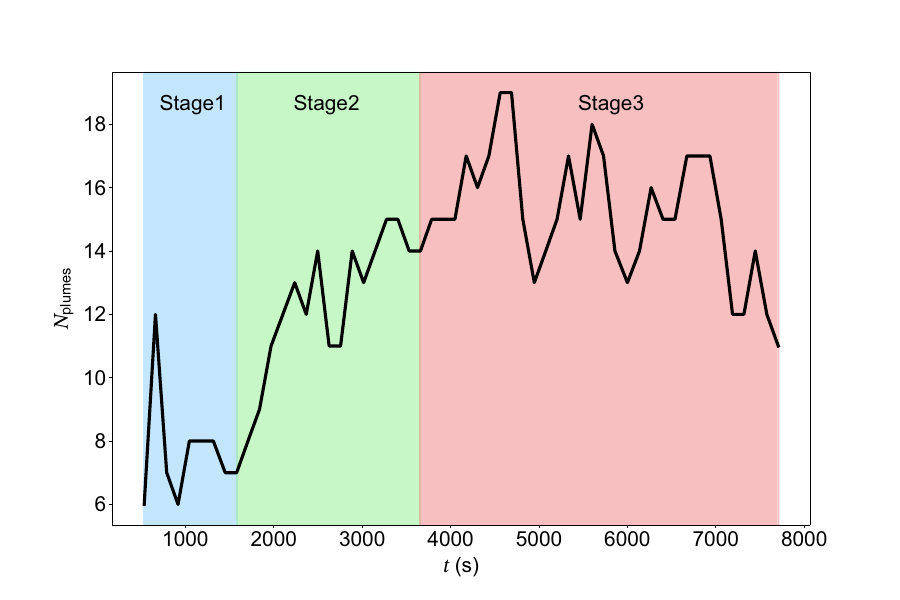}}
  \caption[The number of plumes $N_\mathrm{plume}$ over time]{The number of plumes $N_\mathrm{plume}$ over time. Stage 1 and 2 more transient and stage 3 is more quasi-stable.}
\label{fig:number}
\end{figure}

We do not attempt to extract any precise timescale from the histograms in Fig. \ref{fig:hist}, partly because not enough plumes are available in the distance-time space and because the results depend on bin-size. Instead, we obtain total $N_\mathrm{plume}$ as a function of time (Fig, \ref{fig:number}) by stacking the identified plume centers over the entire SSPIV domain. Visual inspection of Fig. \ref{fig:number} suggests there are three distinct evolutionary stages. In the initiation stage (stage 1, till 1580 s, Epoch 11), $N_\mathrm{plume}$ is on average very low  ($\sim7$), with highs up to 12. The initiation stage is followed by a transient spin up phase (stage 2) which ends at 3659 s (Epoch 27). During stage 2, $N_\mathrm{plume}$ gradually increases to $\sim14$, twice as high as stage 1. The final stage 3 lasts for the rest of the experiment. $N_\mathrm{plume}$ in stage 3 is quasi-stable and ranges from 13 to 19. $N_\mathrm{plume}$ over time provides a handle for different timescales  and corresponding plume dynamics (cf. section \ref{sec:timescales}).

\subsection{Lengthscales}
In this section, we will focus on the lengthscales associated with inter-plume distances, and single plumes.

\subsubsection{Spacing}
\label{sec:spacing}
\begin{figure}
  \centerline{\includegraphics[width=\linewidth]{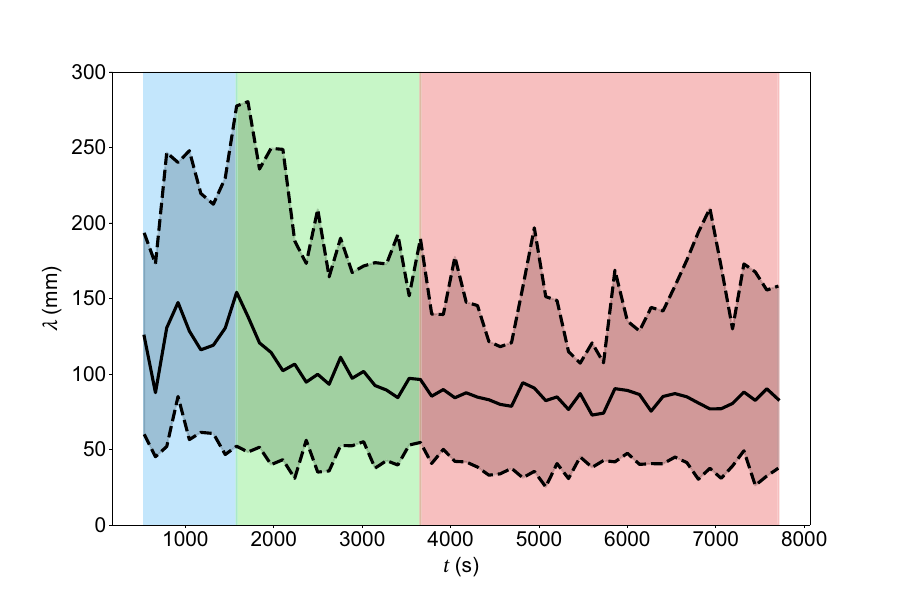}}
  \caption{Plume spacing over time. Average spacing $\Bar{\lambda}$ (solid black line), gray bounds are the minimum and maximum spacing (dashed lines). The range of plume spacing $\lambda$ is encompassed by the shaded region. $\Bar{\lambda}$ is $123.1\pm16.0$ mm ($1\sigma$) in stage 1 (blue shaded region) $106.0\pm17.9$ mm in stage 2 (green), and $83.6\pm5.3$ in stage 3 (red).}
\label{fig:spacing_range}
\end{figure}

The plume spacing, $\lambda$, can also be acquired over time, and a variety of different inter-plume distances can be found. The average plume spacing $\Bar{\lambda}$ (Fig. \ref{fig:spacing_range}) is $123.1\pm16.0$ mm in stage 1. It gradually decays to $106.0\pm17.9$ mm in stage 2, and reaches a quasi-stable spacing of $83.6\pm5.3$ mm in stage 3. Minima (or maxima) in plume spacing follow the same trend, but, as expected, exhibit  stronger (or weaker) fluctuations compared to the average (Fig. \ref{fig:spacing_range}). The minimum spacing is $58.4\pm11.7$ in stage 1, decreasing to $38.3\pm6.1$ in stage 3. The maximum spacing, on the other hand, can be as large as the tank height in stage 1 ($220.3\pm24.9$ mm), and decreases to about half of the tank height ($149.2\pm25.9$ mm) in stage 3. 

\begin{figure*}
  \centerline{\includegraphics[width=\linewidth]{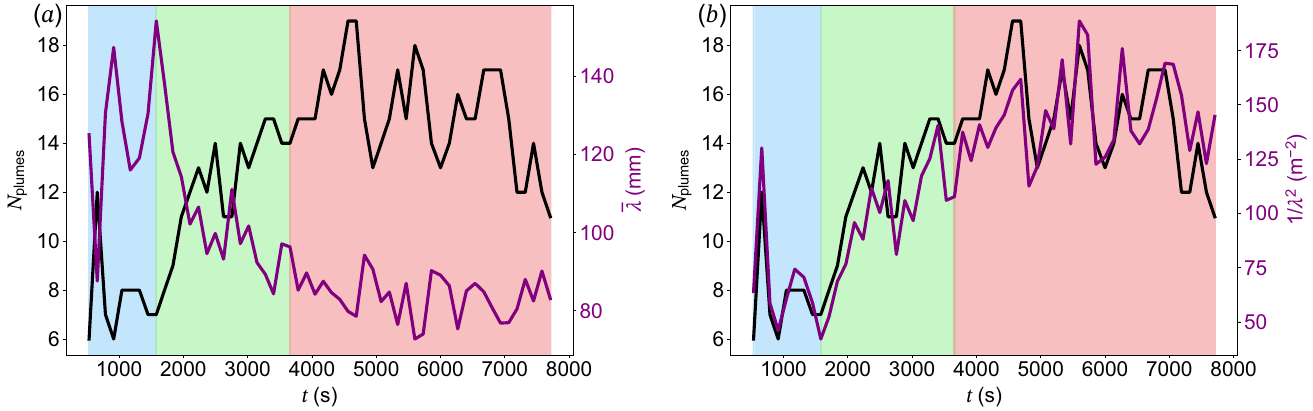}}
  \caption[The number of plumes $N_\mathrm{plume}$ versus the average plume spacing $\Bar{\lambda}$]{The number of plumes $N_\mathrm{plume}$ (black line) versus the average plume spacing $\Bar{\lambda}$ (purple line) in (\textit{a}), and $N_\mathrm{plume}$ versus $1/\Bar{\lambda}^2$ in (\textit{b}) over time.}
\label{fig:spacing}
\end{figure*}

As a function of time, $\Bar{\lambda}^2$ is inversely proportional to $N_\mathrm{plume}$ (Fig. \ref{fig:spacing}). Not surprisingly, as the number of plume increases in the tank, the average spacing decreases (Fig. \ref{fig:spacing}). The average $\Bar{\lambda}^2 N_\mathrm{plume} = 0.1144\pm0.0184 $ m$^2$, remains nearly constant and comparable to the total area of the thermal boundary layer (TBL) supplying the plumes, i.e., the tank bottom ($0.1640$~m$^2$).


\subsubsection{Width of the plume head and stem}
\label{sec:thickness}
Based on plume optical distortion and uneven tracer distributions \cite{bao2024self}, we have estimated the size of the plume head and stem for several representative stages of the experiment.

We find the typical plume head diameter is $\sim60$ mm at mid-tank height for the first batch of plumes ($t=704$ s), which becomes $\sim40$ mm for both in the middle (4091 s) and end (7587 s) of the experiment. The size of the plume head is 18 mm when it has just initiated above the TBL at the beginning of the experiment (at 576 s, from which we estimate the TBL thickness), and grows to as large as $\sim$110 mm when the first batch of plumes hit the surface. Towards the end of the experiment (e.g., at 7587 s), the head size is $\sim65$ mm near the surface \cite{bao2024self}.

The thickness of the plume stem based on the material interface across different plumes is $\delta_\mathrm{m}=6\pm1$ mm, the bottom TBL thickness is $\delta\simeq7.5$ mm comparable to each other \cite<cf. Section 4.4 in >{bao2025piv}.  For the rest of this manuscript we use $\delta$ to refer to both. We estimate the thickness of the top TBL following \cite{Howard1966,Manga1999}:
\begin{equation}
  \frac{\delta_\mathrm{top}}{\delta_\mathrm{bottom}} = 
  \left(\frac{Ra_\mathrm{bottom}}{Ra_\mathrm{top}}\right) ^\frac{1}{3} =
  \left(\frac{\eta(T=25 ^\circ \mathrm{ C})}{\eta(T=80 ^\circ \mathrm{ C})}\right) ^\frac{1}{3} = 4.
  \label{eqn:delta_top}
\end{equation}
to be 24 mm for $\delta=6$ mm, and 30 mm for  $\delta=7.5$ mm.

For clarity, we add redundant subscripts. so that $\delta_\mathrm{bottom}=\delta$, $Ra_\mathrm{top} = Ra$ (the global $Ra$ defined using the top viscosity $\eta(T=25$ $^\circ\mathrm{C})$), $Ra_\mathrm{bottom} = Ra_\delta$ (the local Ra using the bottom viscosity $\eta(T=80$ $^\circ\mathrm{C})$). The distance between two TBLs is $275-6-24=245$ mm for $\delta=6$ mm, and 237.5 mm for $\delta=7.5$ mm. The difference between the two measures is small, so we can use either.  For the inter-TBL distance analysis we use 245 mm but we explore the full range of $\delta=6$ to 7.5 mm for the theoretical analysis of the bottom TBL in \ref{app:TBL}.


\subsection{Timescales}
\label{sec:timescales}
The timescales of plume dynamics tell us about their evolution during their life cycle. Extracting the timescales from our data will eventually provide the speed at which plume material is entrained and transported.

\begin{figure}
  \centerline{\includegraphics[width=\linewidth]{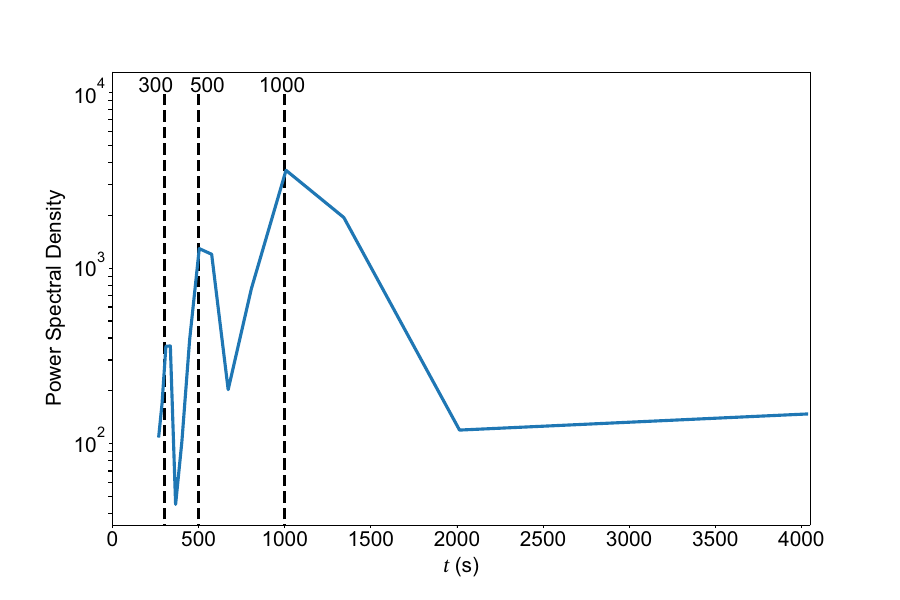}}
  \caption[The power spectrum density at different period for $N_\mathrm{plume}$ at stage 3]{The power spectrum density at different periods for $N_\mathrm{plume}$ at stage 3. Three peaks are around 300 s, 500 s and 1000 s. The $\sim 1000$ s peak is interpreted as a local plume cycle involving conductive rebuilding of the basal TBL, plume initiation, and local depletion, and subsequent replenishment before the next instability.}
\label{fig:PSD}
\end{figure}

We examine the time needed for the plumes to initiate, to rise, to interact with each other, and to finish a full cycle. We focus on the quasi-stable stage 3 with higher $N_\mathrm{plume}$, and calculate its Power Spectral Density (PSD) (Fig. \ref{fig:PSD}). The periods discussed below are therefore characteristic of stage 3, rather than fixed recurrence times for the whole experiment. There are three clear peaks in the PSD -- 300, 500 and 1000 s -- with increasing PSD as the period increases. These peaks can be attributed to differing plume behavior and evolution. 

\begin{figure*}
  \centerline{\includegraphics[width=\linewidth]{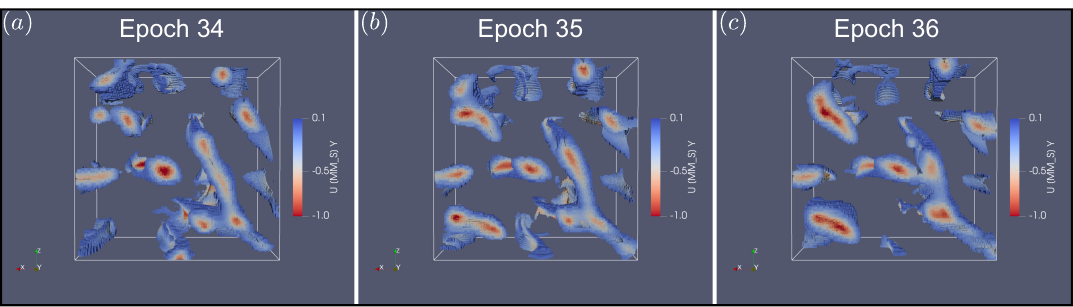}}
  \caption[Bottom to top view of the plume clusters colored by vertical velocity $U_y$]{View from the bottom to the top of the tank of the plume clusters colored by vertical velocity $U_y$. Negative values (red) denote upwellings). The epoch comparison from right to left shows the interaction of upwelling centers. (\textit{a}) Epoch 34, 4561 s.  (\textit{b}) Epoch 35, 4689 s. (\textit{c}) Epoch 36, 4817 s.}
\label{fig:velocity_center}
\end{figure*}

The 300 s period coincides with the timescale of inter-plume interaction. A typical example is shown in Fig. \ref{fig:velocity_center}. Multiple pairs of plumes are connected at the beginning, and merge or almost merge from Epoch 34 to Epoch 36, resulting in lower $N_\mathrm{plume}$. The plumes merge within two epochs ($\sim 260s$), a comparable time interval to the 300 s timescale. The approaching speed is about 0.2 mm/s, resulting in 60 mm of motion in  300 s. Given our 130 s time resolution and a half-epoch uncertainty of 65 s, the 300 s period seems to be clearly associated with the inter-plume interaction timescale (260 s). More details about plume merging will be described in section \ref{sec:merge}. Another example of inter-plume interaction with a similar period will be introduced in section \ref{sec:detach}, where we show an example of the detachment of a plume head after a merging event.

The 500 s period can be linked to plume initiation. From the synchronized velocity fields (cf. Movie S1), the first batch of plumes appears in Epoch 3 ($t=534$ s). By directly examining the pre-synchronization velocity vectors in each plane, i.e., the raw PIV velocity vectors before temporal interpolation \cite{bao2025piv}, we narrow the first plume initiation time to $500\pm5$ s. We therefore take $\tau_{\mathrm{TBL}}\simeq500$ s as the
observed plume-initiation time. This is the amount of time necessary for the boundary layer to develop and become unstable ($\tau_{TBL}$, \citeA{davaille1994onset,Davaille2005}). 

\begin{figure*}
  \centerline{\includegraphics[width=\linewidth]{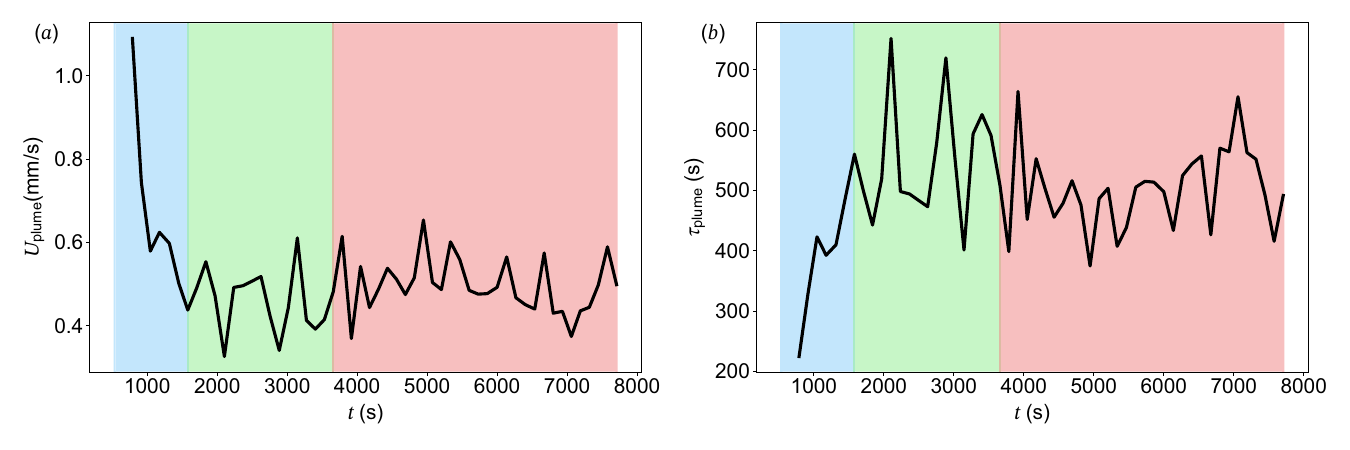}}
  \caption[The characteristic plume rising speed $U_\mathrm{plume}$ and traveling time $\tau_\mathrm{plume}$]{Characteristic plume rising speed $U_\mathrm{plume}$  as a function of time. Positive values denote upwellings. The characteristic speed is measured as the mid-tank average upwelling velocity within the plumes (\textit{a}). Corresponding plume travel time $\tau_\mathrm{plume}$ between the two boundary layers (\textit{b}).}
\label{fig:rising}
\end{figure*}

We also find that the plume travel time $\tau_\mathrm{plume}$ from the bottom to the top TBL matches the 500 s period. Fig. \ref{fig:rising} shows the characteristic rising speed of plumes, as well as the corresponding travel time between the two TBLs. Inspection of the time-dependent FTLE field verifies these values (Movie S2). Although the first batch of plumes can rise as fast as 1 mm/s in stage 1, $U_\mathrm{plume}$ (in this case an average over the stage) decreases dramatically in stage 2 to half the speed (0.5 mm/s) and remains so in stage 3 (Fig. \ref{fig:rising}a). The travel time, as short as $\sim250$ s in stage 1 and as long as $\sim500$ s in stage 3 (Fig. \ref{fig:rising}b), maybe related to the heat supply (cf. Sec. \ref{sec:compare_single}). As the PSD peaks are determined in stage 3, it is notable the 500 s peak matches the independently determined plume travel time $\tau_\mathrm{plume}$.

The last 1000 s period peak in the PSD corresponds to a full local plume cycle, i.e., $\tau_{\mathrm{total}}=\tau_{\mathrm{plume}}+\tau_{\mathrm{TBL}}\approx500+500=1000$ s in stage 3. We interpret this cycle as follows. Heat conduction first rebuilds the basal TBL until it reaches a critical thickness $\delta$ and becomes unstable again. A plume then initiates and drains part of the local thermal anomaly from the bottom TBL into the rising head and stem (cf. \citeA{davaille1994onset,Davaille2005}). Continued basal heating subsequently replenishes the locally depleted boundary layer while the plume travels toward the upper TBL. Thus, $\tau_{\mathrm{TBL}}$ represents the local bottom-boundary-layer rebuilding and instability time, whereas $\tau_{\mathrm{plume}}$ represents the plume travel time between the two TBLs. The 1000 s peak therefore reflects the repeated local depletion--replenishment cycle of the basal TBL coupled to plume ascent, rather than a return of the entire transient system to its original state.

Besides the three periods, longer periods may exist beyond 1000 s seconds (e.g., Fig. \ref{fig:hist}c), perhaps related to downwellings \cite{schaeffer2001interaction}, but cannot be robustly resolved with the time-span covered in the experiment, especially from just stage 3.

We find that for any individual time snapshot, plumes tend to interact with each other rather than being isolated (there are no more than $\sim30\%$ isolated plumes in stage 1 and no more than $\sim60\%$ in latter stages). We have also estimated for how long isolated plumes survive and find that they can last at least 9 epochs or $\sim 1200$ seconds (e.g., cluster 4 from epoch 25, cf. Movie S2).

\section{Results: Dynamical Evolution of Plume and Cluster Morphology}
\label{sec:plume_behaviors}
\begin{figure*}
  \centerline{\includegraphics[width=\linewidth]{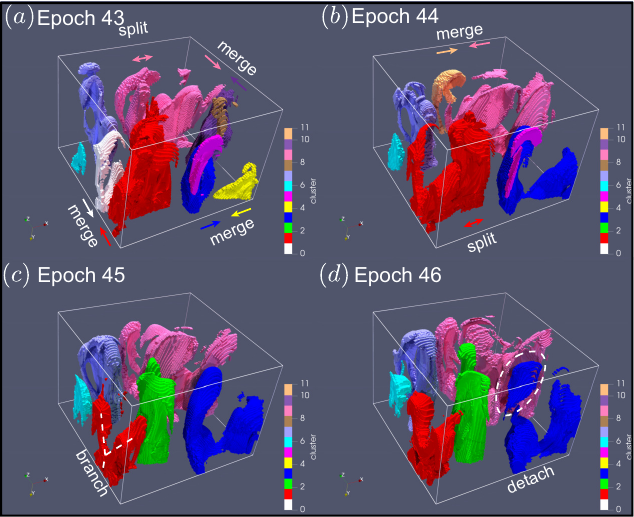}}
  \caption[The dynamics of plume clusters, including splitting, merging, branching and detachment]{The dynamics of plume clusters, including splitting, merging, branching and detachment. Cluster numbers are modified to become ids to be consistent across panels. 
  (\textit{a}) Epoch 43, 5730 s. (\textit{b}) Epoch 44, 5859 s. (\textit{c}) Epoch 45, 6000 s. (\textit{d}) Epoch 46, 6137 s.}
\label{fig:plume_dynamics}
\end{figure*}

\begin{figure*}
  \centerline{\includegraphics[width=\linewidth]{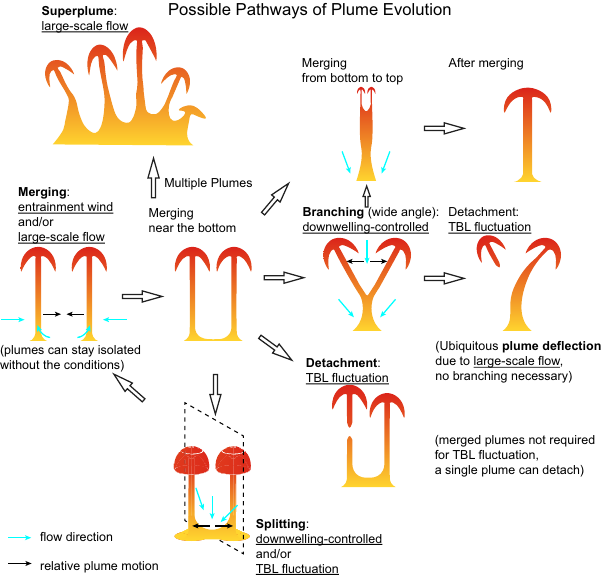}}
  \caption[A zoo of plume dynamical behaviors, possible causes and pathways discussed in this paper]{A zoo of plume dynamical behaviors (bold text), possible causes (underlined text) and pathways (hollow arrows) discussed in this paper. The splitting is shown in 3D, with the blue downwellings in the same plane and perpendicular to the relative plume motion, while others are plotted in 2-D. The large-scale flow in this cartoon emphasizes the lateral component that promotes the shifting of plumes. Note that TBL fluctuation (including the development of the life cycle of a single plume, see section \ref{sec:timescales}) is enough for the detachment of plume head, which does not require merging of plumes. Similarly, large-scale flow is enough for plume deflection, no branching or detachment is needed.}
\label{fig:plume_dynamics_cartoon}
\end{figure*}

The morphology of plumes, their dynamical interactions, and their coupling with the background flow provide key insights into how plume material is distributed and into the signals observed at hotspots and their source regions. Examination of identified plume clusters at different times (e.g., Fig. \ref{fig:plume_dynamics}) reveals rich dynamical behaviors, including merging, superplume formation, splitting, branching, and detachment, arising from interactions with the multi-scale flow (Fig. \ref{fig:plume_dynamics_cartoon}). The robustness of these behaviors is confirmed against the raw velocity field prior to synchronization. These time-resolved observations quantitatively confirm the qualitative plume-interaction regimes in previous studies  \cite<e.g.,>{scott1986observations,Olson1988,Moses1993,schaeffer2001interaction,Gonnermann2004}. Figure \ref{fig:plume_dynamics} shows four consecutive epochs (43–46) containing many of these behaviors, with cluster numbers manually adjusted for consistency across epochs. We describe in detail below each of the identified dynamical behaviors observed in the experiment. We use these examples to document possible plume evolutionary pathways rather than to estimate event probabilities; a robust occurrence probability would require automated tracking and classification of plume-cluster topology through time, as well as  experiments under similar conditions, and will be the subject of future work.

\subsection{Merging}
\label{sec:merge}
The most common behavior is the merging of plumes. For example, clusters 0 and 1, clusters 3 and 4, and clusters 9 and 11 in Fig. \ref{fig:plume_dynamics}a, b illustrate this behavior. Before their merging, the cluster pairs may be already close to each other (e.g., $\sim35$ mm between the edge of the two clusters). Note that the lateral flow surrounding the stem of a single plume is inward flow (flow towards the plume, or ``entrainment wind", \citeA{Schubert2004}). A plume at the edge of a cluster is also associated with this entrainment wind, which is typically $\sim0.2$ mm/s, and up to 0.5 mm/s locally near the base for the strongest, fastest plumes. As a result, two closely located clusters tend to merge, if there is no local downwelling between the clusters. 

\begin{figure*}
  \centerline{\includegraphics[width=\linewidth]{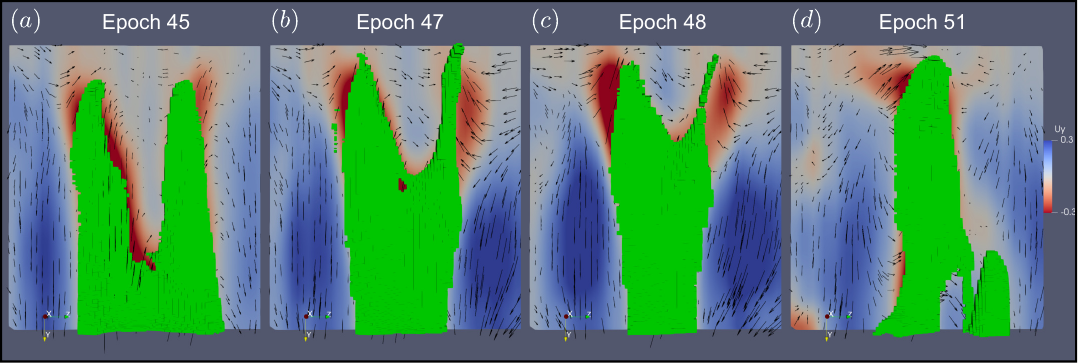}}
  \caption[The complete merging from bottom to top of two plumes in cluster 2 since Epoch 45]{The complete merging from bottom to top of two plumes in cluster 2 starting at Epoch 45. 
  The downwellings near the splitting plume cluster 1 in Epoch 44 as in Fig. \ref{fig:plume_dynamics}b. A $U_y$ (in mm/s) cross section ($x=110$ mm) are plotted with black arrows showing the 3D velocity vector. Only a negligible or small downwelling is between the two plumes. (\textit{a}) Epoch 45, 6000 s.  (\textit{b}) Epoch 47, 6268 s. (\textit{c}) Epoch 48, 6403 s. (\textit{d}) Epoch 50, 6673 s, a new plume also emerges at the z positive direction.}
\label{fig:full_merge}
\end{figure*}

A full-merging event of two plumes within cluster 2 (Fig. \ref{fig:plume_dynamics}c, d) is shown with a zoomed-in view in Fig. \ref{fig:full_merge}. The two plumes merge gradually from the bottom to the head, with the two stems becoming smaller and narrower over time (also see Fig. \ref{fig:plume_dynamics_cartoon}).

This merging could be aided by the large-scale flow, if the two clusters are surrounded by downwelling flow. For example, downwellings may push two distant clusters (e.g. 60 mm distance, cf. section \ref{sec:timescales}) to be closer within an epoch or two (see Fig. \ref{fig:velocity_center} and Movie S1,2). 

It is worth noting that while the upper part of two individual plumes can sometimes be connected (usually within a cluster, e.g., cluster 1 in Epoch 43, Fig. \ref{fig:plume_dynamics}a, Movie S2) in our experiment, we could not clearly resolve two isolated plumes merging from the upper part instead of the bottom as seen by \citeA{Moses1993}. However, this does not exclude the existence of this alternative merging scenario, because it is possible that our limited temporal resolution causes the FTLE ridges \citeA{bao2025piv} to widen enough to obscure this merging mode. 

Similar merging of neighboring plume-like structures were observed by \cite{whitehead1970thermal} for cold jets descending through a radiatively heated upper unstable layer into a stably stratified fluid. Their thermal configuration differs from our basally heated system, but their observations provide an early analogue for jet/plume merging. In our experiment, gradual warming of the bulk fluid introduces an internal-heating-like tendency in the transient mean thermal budget, but the heat is still supplied through the basal boundary and the observed structures remain bottom-rooted hot upwellings; this warming primarily weakens the effective thermal/viscosity contrast and modifies the boundary-layer forcing through time.

\subsection{Superplumes}
\label{sec:superplume}
\begin{figure*}
  \centerline{\includegraphics[width=\linewidth]{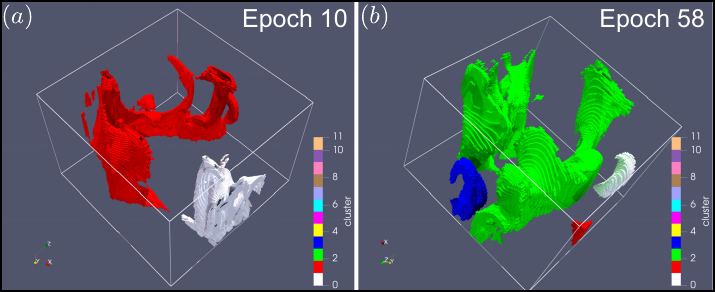}}
  \caption[Examples of superplumes shown with plume clusters]{Examples of superplumes shown with plume clusters.  (\textit{a}) Epoch 10, 1451 s.  (\textit{b}) Epoch 58, 7707 s.}
\label{fig:super_plume}
\end{figure*}

Plumes can shift as they are swept by the large-scale flow. Multiple plumes can be pushed together (e.g., 3 or more), to form a ``superplume" (Fig. \ref{fig:plume_dynamics_cartoon}). The superplume can be created either early in the convection when the flow is more transient (e.g., Epoch 10, Fig. \ref{fig:super_plume}a), or very late when the flow is quasi-stable (e.g., Epoch 58, Fig. \ref{fig:super_plume}b). Not surprisingly, when a superplume exists, it accounts for a large fraction of the total buoyancy flux. We can therefore use the number of plumes as a proxy for the buoyancy flux, i.e., the number of plumes inside the superplume versus those outside. This is a good proxy, as we do not measure temperature or heat flux within the tank and cannot independently compute the buoyancy flux. In the two epochs were the superplume is present shown in Fig. \ref{fig:super_plume} , it accounts for  $>70\%$ of the buoyancy flux at that time. However, this fraction can be as low as $23\%$ (superplume in Epoch 27, see Movie S1,2). The presence and heat flux fraction of the superplume might be dependent on various parameters, including the aspect ratio of the box ($\sim$1.5 in this experiment), because the aspect ratio and side walls could influence the geometry of the large-scale return flow that can sweep plumes together into a composite cluster. In the present experiment, we could not identify a single persistent, coherent large-scale downwelling cell throughout the run (Movie S1), so the dependence of superplume formation and flux partitioning on aspect ratio and return-flow geometry remains a topic for future experiments.

The identification of a superplume is mainly based on whether several plumes are connected by the same FTLE ridge, and largely controlled by the FTLE threshold. We have chosen a conservatively low FTLE threshold to pick up as many individual plumes as possible. Increasing the FTLE threshold may change the extent of the superplume, but at a cost of missing smaller, weaker plumes. Note that the present FTLE threshold 0.005 s$^{-1}$ can effectively distinguish the background and the plumes, as we explained in the companion paper \cite<cf. Appendix D, Fig. D4 in >{bao2025piv}.

\subsection{Splitting}
\label{sec:split}
\begin{figure*}
  \centerline{\includegraphics[width=\linewidth]{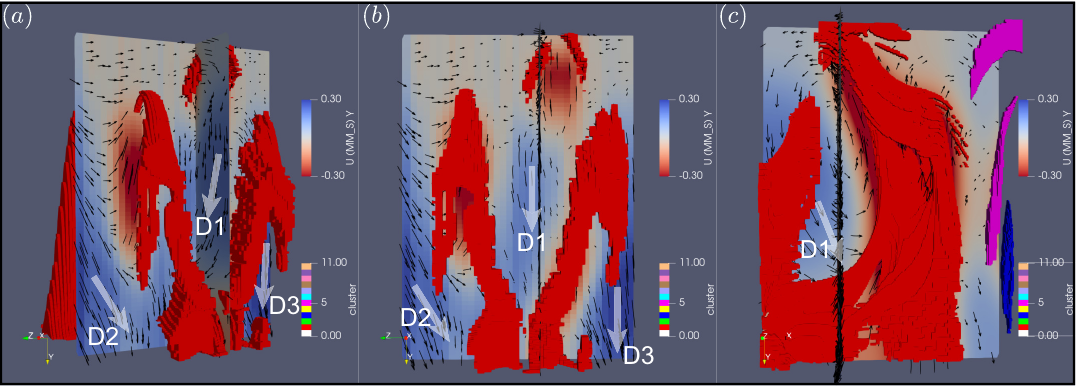}}
  \caption[The downwellings near the splitting plume cluster 1 in Epoch 44 as in Fig. \ref{fig:plume_dynamics}b]{The downwellings near the splitting plume cluster 1 in Epoch 44 as in Fig. \ref{fig:plume_dynamics}b. Two $U_y$ (in mm/s) cross sections ($x=40$ mm,$z=-270$ mm) are plotted with black arrows showing the 3D velocity vector. 3 regions of downwellings denoted as D1, D2 and D3, are highlighted with white arrows. Three different view angles from $z+$/$x-$, $x-$ and $z-$ are plotted in (\textit{a}), (\textit{b}) and (\textit{c}), respectively.}
\label{fig:donwelling3}
\end{figure*}

Downwellings do not always promote the merging of clusters. During the merging of cluster 0 and 1 in Epoch 43 (Fig. \ref{fig:plume_dynamics}a), a downwelling region starts to develop between the clusters (denoted as D1). Downwelling D1 is limited to the upper half of the tank in Epoch 43. Before the downwelling can further develop, cluster 0 and 1 have merged from the bottom, leaving a gap between the upper half of the two clusters (Fig. \ref{fig:donwelling3}). Downwelling D1 continues to develop through this gap towards the x positive direction in Epoch 44. Meanwhile, two additional downwellings (D2 and D3, Fig. \ref{fig:donwelling3}) from z positive and negative directions ``attack" the neck connecting the two parts of the plume cluster along x. The two parts then fully separate from each other, becoming clusters 1 and 2 in Epoch 45 (Fig. \ref{fig:plume_dynamics}b, c). The three downwellings D1, D2 and D3 may limit the growth of the upwelling material at the neck. And at the same time, the TBL at the neck may drain out due to the previous merging event. That is, more than one plume stem within the connected root is now entraining material from the same basal area of the TBL, even though they were originally sampling different regions of the TBL. The draining of the TBL and the downwelling flow lead to the splitting of cluster 1 in Epoch 44 to clusters 1 and 2 in Epoch 45 (Fig. \ref{fig:plume_dynamics_cartoon}).

Clusters 9 and 11 are another example of splitting and re-merging in Epochs 43 and 44 (Fig. \ref{fig:plume_dynamics}a, b). In this case, given the absence of a downwelling (Movie S1,2), this splitting is primarily due to TBL fluctuations alone. That is, the plumes stems that share the same root temporarily drain out the TBL between clusters 9 and 11; later, the TBL grows back. 

\subsection{Branching}
\label{sec:branch}
After the splitting, the new cluster 1 shows a branching structure resulting from the merging (cluster 0 and 1, Epoch 43) and splitting (cluster 1 and 2, Epoch 45), i.e., two heads from the same root (dashed line in Fig. \ref{fig:plume_dynamics}c). This branching is unique, because it has two strongly tilted plume stems upon merging. Such wide-angle branching is controlled from downwellings D1, D2 and D3, especially D1 which separates the two stems.

By contrast, Fig. \ref{fig:full_merge} shows a typical example of two plumes fully merging. Before the full merging, the two plumes are connected at the base, but their stems are nearly parallel. This ``narrow-angle" branching is due to the absence of a strong downwelling like D1 (Fig. \ref{fig:plume_dynamics_cartoon}).

In the case of the ``wider-angle" branching in cluster 1 (Fig. \ref{fig:plume_dynamics}c), the two plume heads can stay away from each other and never merge, even when the merging develops further away from the TBL. This is partly because one plume branch detaches from the stem (see Epoch 44-48 Movie S2). We will demonstrate detachment behavior with another example in the next section.

\subsection{Detachment and pulsing}
\label{sec:detach}
\begin{figure*}
  \centerline{\includegraphics[width=\linewidth]{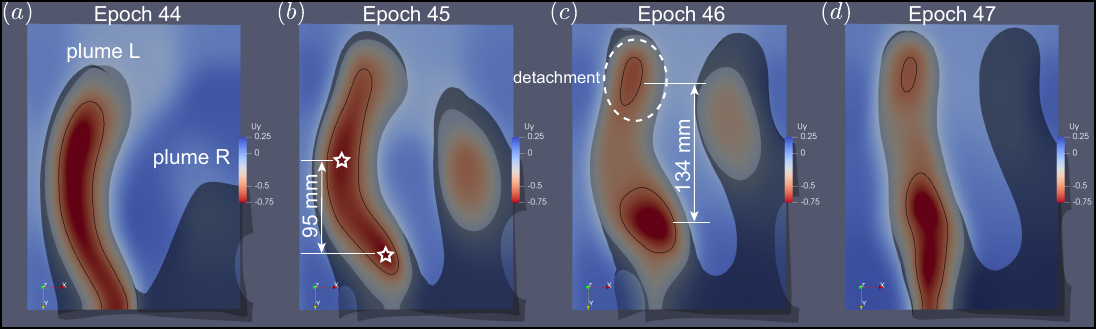}}
  \caption[Vertical velocity of plume cluster 3 with head detachment as in Fig. \ref{fig:plume_dynamics}]{Vertical velocity of plume cluster 3 with head detachment as in Fig. \ref{fig:plume_dynamics}. $U_y$ (in mm/s, negative for upwelling) is plotted in the background plane ($z=-275$ mm), with the 2-d black contour lines showing $U_y=-0.5$ mm/s in the plane, and the 3D black contour surfaces showing $U_y=-0.12$ mm/s, so the contour surfaces intersect at the plane at $U_y=-0.12$ mm/s as well (enclosing the bright area inside the darker area). (\textit{a}) Epoch 44, 5859 s. (\textit{b}) Epoch 45, 6000 s. The velocity centers are shown with stars. (\textit{c}) Epoch 46, 6137 s. (\textit{c}) Epoch 47, 6268 s.}
\label{fig:Uy_detach}
\end{figure*}

In cluster 3 there is a clear plume head detachment event (Fig. \ref{fig:plume_dynamics}d, \ref{fig:plume_dynamics_cartoon}). The head disappears (identified by the absence of the FTLE ridge in the upwelling domain) in Epoch 47, while the stem and root remain stable (Movie S2). Examination of $U_y$ suggests that the material supply from the stem does not provide enough buoyancy flux to maintain the the head (Fig. \ref{fig:Uy_detach}).

Specifically, we observe pulsing velocities in the plume, which manifest as alternating regions of strong upwelling (Fig. \ref{fig:Uy_detach}) rather than a steady flow. Before the detachment in Epoch 44, plume cluster 3 (plume L) and 4 (plume R) have just merged from the bottom (Fig. \ref{fig:plume_dynamics}b). At this time, plume L shows a relatively smooth and coherent $U_y$ from top to bottom (Fig. \ref{fig:Uy_detach}a). In the next step (Epoch 45), however, two distinct $|U_y|$ local maxima appear above the TBL in plume L, separated vertically by 95 mm, signaling that the head is about to detach (Fig. \ref{fig:Uy_detach}b). By Epoch 46, after the head has detached from the stem, the two $|U_y|$ maxima are farther apart (134 mm, Fig. \ref{fig:Uy_detach}c), indicating a pulse-like separation of upwelling zones. A simple scaling, dividing the vertical distance between maxima by the mean upwelling speed $U_\mathrm{plume}=0.5$ mm/s, gives a pulsing timescale of 190–268 s, suggesting than intra-plume timescales are, not surprisingly, shorter than the 300 s inter-plume interaction time scale identified above. In Epoch 47, after detachment, a new strong upwelling region connected to the TBL emerges (Fig. \ref{fig:Uy_detach}d), showing the recovery of buoyancy supply from the TBL.

This pulsing behavior is likely related to fluctuations (draining and regrowth) in the TBL, either right beneath plume L, or the neighboring plume R, or both. The merging of plumes L and R may have redistributed and modulated the supply of buoyancy necessary to sustain the individual plumes. This pulsing timescale (190 to 268 s) is about half of the plume initiation timescale ($\sim500$ s), and comparable to the growth to at least $\sim70\%$ of the bottom TBL $\delta$ (growing timescale $\tau_\mathrm{grow}\propto \delta^2$) .

It is interesting to note that plume L is also an example of a deflected plume. The deflection is a direct result of the interaction of the ambient large-scale flow with the plume. Most of the plumes we observed in the experiment ($>70\%$) are vertical or sub-vertical (e.g. Fig. \ref{fig:velocity_ftle_cluster}), but deflections are not uncommon (e.g., Fig. \ref{fig:donwelling3}c). The ubiquitous influence of the large-scale flow on individual plumes (i.e., not inter-plume interactions) also includes the shifting of plumes leading to merging and superplume clusters (cf. section \ref{sec:merge}, \ref{sec:superplume}). 

Plume deflection can be also caused by processes not included in our experiment, like plate-driven flow (e.g., \citeA{skilbeck1978formation,whitehead1982instabilities}), or the presence of mid-mantle low viscosity layers (e.g., \citeA{yuen1998mesoscale}).

\section{Discussion}
\label{sec:discuss}
Our 4D velocity measurements, together with the end-to-end methodology described in \citeA{bao2025piv}, enabled a detailed quantitative view of plume dynamics—including their initiation, development, morphology, number, spatiotemporal evolution, and interactions among plumes and with the background flow. These results not only allow direct validation and comparison with previous studies but also yield new insights made possible by the rich information we are able to extract from a single experiment. In particular, we are able to propose evolutionary pathways for plumes beyond static morphologies, which will be important to analyze material entrainment and enable surface-to-source mapping and interpretation of advanced seismic imaging. The fully characterized plumes in Rayleigh-B\'enard flow in our experiments let us ultimately address outstanding questions for plume flow in planetary mantles. These include: the location of plumes, the material they entrain, the reservoirs that can be sampled by the plumes, as well as the chemical and dynamical relationship between plumes and the lowermost mantle.

\subsection{Comparison with theory and previous experiments}
\subsubsection{Comparison with single thermal plume experiments}
\label{sec:compare_single}
In this section, we compare our basal heating experiment with previous scalings \cite<e.g.,>{Batchelor1954,olson1993structure,Moses1993,Davaille2011} and the spot heating experiment in \citeA{Cagney2015}, in terms of plume rising speeds, buoyancy supply and entrainment. Despite the similarities between the two experimental studies, the differences suggest it is necessary to use basal heating to capture plume characteristics more accurately (such as the rising speed), especially for application to the Earth and other planetary mantles. The scaling to the Earth will be further discussed in section \ref{sec:discuss_speed}.

\citeA{Cagney2015} heated corn syrup with almost identical properties to the fluid in this study in a cubic tank with internal dimensions of 265 mm, and a focused heating source 20 mm in diameter, to achieve a $\Delta T$ of 55 $^\circ$C and a similar tank $Ra$ ($1.4\times10^6$).

Our plume rising velocity is about 4 to 8 times (Fig. \ref{fig:rising}) as high as those in the single thermal plume experiment ($\sim0.12$ mm/s) of \citeA{Cagney2015}. Both measurements were taken during the stage in the plume life-cycle where the rising velocity was constant, in other words, neither during the acceleration stage when the plume initiates, nor during deceleration when the plume feels or has hit the top boundary (cf. Fig. 11 in \citeA{Cagney2015}). The only deviation in our experiment is at the end of transient stage 1, when all plumes decelerate, and $U_\mathrm{plume}$ is underestimated. 

The difference in rising velocity is unlikely to be the result of the somewhat larger domain in our experiment. In other words it is not a wall effect. The velocity deficit from the viscous drag from the presence of the wall in low $Re$ convection can be approximated by a rising sphere in a square tube \cite{Happel1974}:

\begin{equation}
  U_\mathrm{rising} =  U_\mathrm{rising}^\infty (1-1.903D/W)
  \label{eqn:wall_effect}
\end{equation}
where $U_\mathrm{rising}$ and $U_\mathrm{rising}^\infty$ are the rising speed within a finite container or in infinite space, respectively. $D$ is the diameter of the sphere and $W$ is the width of the square container. If we take $D$ from the width of a typical plume head at mid-tank height, in our experiment \cite{bao2025piv} $D\sim60$ mm and $W=405$mm. In \citeA{Cagney2015}, $D\sim40$ mm, and $W=265$ mm. The relative velocity deficits are comparable ($1.903D/W$): $28\%$ in this study and $29\%$ for \citeA{Cagney2015}. With more than one plume in our experiment, the corresponding deficits could be even larger. 

Rather than wall effects, the difference in plume material supply is the more likely cause of the faster rising velocity. In \citeA{Cagney2015}, the tank is only heated from a focused region (a circular copper heater 20 mm in diameter) and a substantial fraction of the bottom boundary layer is cold. In our tank there is uniform basal heating, and the entire bottom TBL is hot. It has been previously suggested that $U_\mathrm{plume} \propto Q^\frac{1}{2}$ \cite{Batchelor1954,olson1993structure,Moses1993,Davaille2011}. Given the similar conditions in \citeA{Cagney2015}, we can further estimate the diameter $D_\mathrm{heat}$ for the effective circular area of heat flux supply for each of our plumes from the heater size. Since $Q\propto D_\mathrm{heat}^2$, $U_\mathrm{plume} \propto D_\mathrm{heat}$.  Given that our $U_\mathrm{plume}$ is  4 to 8 times that of \citeA{Cagney2015}, we estimate $D_\mathrm{heat}$ to be 80-160 mm. This is comparable to our average plume spacing $\Bar{\lambda}\sim100$ mm (cf. section \ref{sec:spacing}). We conclude that our faster speeds are due to more heat flux $Q$ and thermal buoyancy being available to the plumes in our experiment.


We can further compare $D_\mathrm{heat}$ against $\Bar{\lambda}$ at different stages in our experiment. In stage 1, $D_\mathrm{heat}$ is $\sim$ 160 mm for the fastest plumes, while $\Bar{\lambda}$ can be as large as 150 mm. $D_\mathrm{heat}$ decreases towards stage 3 to $\sim$ 80 mm, which is identical to $\Bar{\lambda}$ in stage 3 (Fig. \ref{fig:spacing}). This suggest that to first order, $D_\mathrm{heat} \sim \Bar{\lambda} \propto N_\mathrm{plume}^{-2}$, which confirms that the entire bottom TBL area provides heat to the plumes. The measured dependence of $N_{plume} \propto \Bar{\lambda}^{-2}$ agrees with viscous-convection scaling expectations \cite{wenzel2004effect,zhong2005dynamics}. The heat supply area diameter $D_\mathrm{heat}$ in stage 3 is comparable to the estimated diameter (78.52 mm) of the feeder region for the single plume experiment with a spot heater in \citeA{Cagney2015}. As the plume stem widths are also similar, it suggests material entrainment for plumes from spot heating and fully basally heated systems are the same. This is not the case for plumes generated by injection \cite{Newsome2011}. Hence, despite the larger head sizes and rising speeds in our experiments, the material entrainment will be similar to the spot heating experiments.


\subsubsection{Plume thickness $\delta$, head diameter $D$, and initiation timescale $\tau_{TBL}$}
\label{sec:discuss_thickness_tau_head}

The plume head and stem size, the bottom TBL thickness, and the plume-initiation timescale are closely linked because as the TBL grows conductively and becomes unstable. Since the image-based plume stem thickness and the independently inferred lower-TBL thickness are comparable (section \ref{sec:thickness}), we use $\delta$ to denote both the TBL and plume-stem thickness in this discussion. The classical boundary-layer instability scaling \cite{Howard1966} can be written either as a timescale estimate,
\begin{equation}
  \tau_{TBL}
  =
  \frac{H^2}{p\kappa}
  \left(
      \frac{Ra_c}{Ra}
  \right)^{2/3},
  \label{eqn:tau_TBL_main}
\end{equation}
or equivalently as a thickness estimate,
\begin{equation}
  \delta
  =
  \sqrt{p\kappa\tau_{TBL}}
  =
  H
  \left(
      \frac{Ra_c}{Ra}
  \right)^{1/3}.
  \label{eqn:delta_main}
\end{equation}
Here $Ra_c$ is the critical Rayleigh number and $p$ is a prefactor whose value depends on the operational definition of the conductive TBL thickness (cf. \ref{app:TBL}). The thickness estimate in equation \eqref{eqn:delta_main} is independent of $p$ once $Ra$ and $Ra_c$ are specified. Using the analytical solution for the no-slip boundary condition from \citeA{Jeffreys1928}, $Ra_c=1708$, also consistent with experimental estimates which range from 1300 to 2000 (e.g. \citeA{Sparrow1970,Manga1999,LeBars2004}). $Ra_c$ should increase by no more than 15\% due to the temperature-dependent viscosity \cite{Stengel1982} even with the max viscosity contrast in this experiment ($\gamma=65$).

Using $Ra_c=1708$ and the second part of Equation \ref{eqn:delta_main}, the cold/global Rayleigh number $ Ra=1.9\times10^6$ gives $\delta\simeq26.5$ mm, whereas the hot-bottom/local Rayleigh number $Ra_{\delta}=1.2\times10^8$ gives $\delta\simeq6.7$ mm. The local-Rayleigh-number estimate is the one consistent with the observed active lower TBL and plume stem thickness, $\delta\simeq6$-$7.5$ mm. The global $Ra$ estimate instead corresponds to a much colder outer conductive penetration depth.

The observed first plume-initiation time gives an independent way to interpret the prefactor $p$. The first plume batch appears at about $t=500$ s, with the bottom temperature increased approximately linearly from room temperature to $80^\circ$C over the first 250 s. The ramp-corrected conductive growth time, $\tau_{eff}$ is therefore:
\begin{equation}
  \tau_{\mathrm{eff}}
  \simeq
  500-\frac{250}{2}
  =
  375~\mathrm{s}.
\end{equation}
Combining $\tau_{\mathrm{eff}}\simeq375$ s with $\delta=6$-$7.5$ mm gives a prefactor
\begin{equation}
  p
  =
  \frac{\delta^2}{\kappa\tau_{\mathrm{eff}}}
  \simeq
  0.9\text{-}1.4.
\end{equation}
For an erfc-style conductive profile, this range corresponds to a mid-temperature isotherm, approximately $\Theta\simeq0.4$--0.5, rather than to the cold outer edge of the conductive profile. The commonly used value $p=\pi$ \cite{Howard1966} has a more specific meaning: it is the thickness of a linear temperature profile  heat-flux-equivalent, or basal-slope-equivalent to the erfc conductive profile. It is therefore not defined originally from the position of a real isotherm. If $p=\pi$ is imposed with the local Rayleigh number in Equation \ref{eqn:delta_main}, the predicted initiation time is shorter than observed; if it is imposed with the global Rayleigh number, both the predicted thickness (Equation \ref{eqn:tau_TBL_main}) and initiation time (second part of Equation \ref{eqn:delta_main}) are too large.

We find similar results with the properties and experimental conditions in \citeA{Davaille2005}.
Applying the same local-Rayleigh-number estimate to their experiment gives a TBL thickness corresponding to a similar nondimensional temperature $\Theta=0.47$ and a similar prefactor $p=1.04$. In contrast, a particular plotted or visible low-temperature isotherm with thermochromic liquid crystals can correspond to a much colder outer front and therefore to a larger apparent TBL thickness. For example, the 24.6 $^\circ$C isotherm used in \citeA{Davaille2005} is equivalent to $\Theta\sim0.1$, while in our experiment an isotherm-defined edge at $\Theta=0.1$ would correspond to a thickness of about 15 mm (compared to 12.7 mm in \citeA{Davaille2005} at the ramp-corrected initiation time, larger than the 6 to 7.5 mm determined from our raw image, or the $\Theta\simeq0.4$--0.5 active plume-root scale. Thus, different reported TBL thicknesses can arise from different operational definitions of the boundary layer. Full discussion and derivation of $p$, the ramp correction, and the mapping between $p$, $\Theta$, $\delta$, and $\tau_{TBL}$ can be found in \ref{app:TBL}.

We next use a localized, TBL Rayleigh-number argument to evaluate how $\delta$ should evolve as the fluid interior becomes progressively warmer. Define
\begin{equation}
  Ra_{\mathrm{TBL}}
  =
  \frac{\rho g\alpha \Delta T_{\mathrm{TBL}}\delta^3}
  {\kappa\eta_{\mathrm{eff}}},
  \label{eqn:Ra_TBL_main}
\end{equation}
where $\Delta T_{\mathrm{TBL}}$ is the temperature drop across the lower TBL and $\eta_{\mathrm{eff}}$ is an effective viscosity for the active TBL. If plume-producing TBL instability occurs at an approximately fixed effective $Ra_{\mathrm{TBL}}$, then
\begin{equation}
  \delta
  \propto
  \left(
      \frac{\eta_{\mathrm{eff}}}
      {\Delta T_{\mathrm{TBL}}}
  \right)^{1/3}.
  \label{eqn:delta_evolution_main}
\end{equation}
As the interior warms, $\Delta T_{\mathrm{TBL}}$ decreases, but the effective viscosity sampled by the lower TBL also decreases because the TBL itself becomes warmer. These two effects compete. Using the measured viscosity law of our working fluid and estimating $\eta_{\mathrm{eff}}$ from the mean temperature across the lower TBL, \ref{app:TBL} shows that the expected relative thickness remains approximately constant over the relevant range of interior temperatures. The basal heat flux nevertheless decreases because
\begin{equation}
  q_{\mathrm{TBL}}
  \sim
  k\frac{\Delta T_{\mathrm{TBL}}}{\delta}.
\end{equation}
The observed decrease in heat supply is therefore expressed mainly in the plume rising velocity and plume-head size, while the TBL/stem thickness remains approximately constant. We observe exactly this behavior (Fig. \ref{fig:rising}).

The plume head provides a separate but related length scale. The source-fed head scaling of \citeA{whitehead1975dynamics}, \ref{app:head} predicts a newly initiated plume-head diameter $\sim 14-21$ mm, with a value near 18 mm for the first batch of plumes in our experiment, consistent with the observed head size just above the lower TBL. We also use our average plume spacing (Fig. \ref{fig:spacing_range}) to estimate the feeding area and following \citeA{whitehead1975dynamics} to estimate the head size for each plume.  Assuming the head is first order spherical, we predict mid-tank head diameters of about 42-60 mm across the observed spacing range. This agrees  well with the observed $\sim60$ mm mid-tank heads in the first batch and the smaller $\sim40$ mm heads in stage 3. The representative plume-head diameter then increases approximately linearly with height during ascent: from 18 mm near initiation to $\sim60$ mm near mid-tank and as large as $\sim110$ mm near the upper boundary for the first batch, and from a lower-end initiation estimate of $\sim15$ mm, to $\sim40$ mm near mid-tank and $\sim65$ mm near the upper boundary in stage 3. This approximate height-dependent growth is consistent with the continued feeding and expansion of starting thermal plumes observed in previous single-plume experiments \cite{Newsome2011,Cagney2016a}. For stage-3 mid-tank heads, $D\simeq40$ mm and $\delta=6$--$7.5$ mm lead to a head to stem ratio $D/\delta\simeq5.3$--6.7, consistent with the head/stem ratio of 5 reported by \citeA{lithgow2001plume} near $Ra\simeq10^8$.

\subsubsection{Plume spacing}
\label{sec:spacing}
Previous experiments suggest the plume spacing to be between 3 to 6 $\delta$ \cite{Sparrow1970,Tamai1984,Asaeda1989,Davaille2002,Jellinek2004,Androvandi2011}, which would correspond to 18 to 36 mm in our experiments. This is much smaller than our average spacing $\Bar{\lambda}$ ($\sim80$ mm in stage 3), but closer to our determined minimum spacing ${\lambda_\mathrm{min}}$ of $38.3\pm6.1$ mm in stage 3, and in agreement with the upper bound of $\lambda_\mathrm{min}/\delta \sim 6$ if we use the lower end of our $\delta=6$ mm, while $\delta=7.5$ mm yields a ratio of 5.1.

It is possible that previous experiments might have underestimated the average spacing as the available statistics were only from a selected vertical PIV cross section, or projected shadowgraphs. By contrast, we are able to systematically evaluate the overall distribution of all plumes in 3D space, and in time. This explains why our spacing/TBL ratio is larger. The spacing in a 2-D vertical projection, cannot be larger than the real spacing in 3D. Hence, measurements on the 2-D cross section are closer to the minimum rather than average spacing when the full 3D space is considered. This reinforces the importance of our full 4D quantitative characterization of the flow for plume statistics and evolution.

We can also try to estimate the spacing from linear stability theory. Consider plumes as hot lighter fluid entering a cold, more viscous, denser fluid on top, in a flow with a no-slip basal boundary as in \citeA{Lister1989,Jellinek2002}. The spacing is given by:

\begin{equation}
  \lambda_\mathrm{lst} = \frac{2\pi}{C}\gamma_\mathrm{interface}^{1/3}\delta
  \label{eqn:spacing}
\end{equation}
where $C$ is a prefactor and $\gamma_\mathrm{interface}$ is the viscosity contrast of the upper fluid over the lower fluid at the interface (the lower fluid in \citeA{Lister1989} now becomes the tank bottom, effectively infinitely dense and viscous). In our case $C=(\frac{3}{2})^{1/3}$,  the coefficient $\frac{2\pi}{C}$ is $5.49$, and $\gamma_\mathrm{interface}\ge1$ near the top of the bottom TBL. We then obtain $\lambda_\mathrm{lst}$ to be at least $33$ mm, consistent with our experimentally-determined ${\lambda_\mathrm{min}}$. We note that \citeA{whitehead1975dynamics} suggested a spacing scaling with a similar form, for a buoyant low-viscosity layer of thickness $\delta$ rising into a more viscous overlying fluid with a no-slip bottom boundary condition:

\begin{equation}
\lambda_\mathrm{wl} = \frac{2\pi}{k_\mathrm{wl}}\epsilon^{1/3}\delta
\simeq 2.92\delta\epsilon^{1/3}
\label{eqn:spacing2}
\end{equation}

where $k_\mathrm{wl}=2.15$ is the fastest-growing wavenumber and $\epsilon$ is the viscosity ratio of the overlying fluid over the buoyant layer. Although the coefficient 2.92 is smaller than 5.49 in Equation \ref{eqn:spacing}, here $\epsilon$ describes the effective viscosity contrast of the two layers instead of that at the interface and absorbs the coefficient difference. Using Equation \ref{eqn:spacing2}, effective $\epsilon$ inferred from the minimum spacing is 19 to 37 for stage 1 with $\delta=7.5$ or $6$ mm, respectively; and 5 to 10 for stage 3. This is below our full viscosity contrast $\gamma=65$ and consistent with the gradually warmer fluid interior over time.

Lastly, we compare our plume spacing with the plume population statistics of the numerical simulation in \citeA{parmentier2000three}. We define the non-dimensional plume spacing as $\lambda'=\lambda/H$, where $H$ is the fluid-layer thickness. \citeA{parmentier2000three} considered isoviscous, internally heated infinite-Prandtl-number convection cooled from above, with an adiabatic bottom thermal boundary, free-slip top and bottom mechanical boundaries, and either symmetric or periodic vertical side boundaries. From their reported plume counts, the inferred non-dimensional plume spacing at $Ra=10^8$ is $\lambda'\approx0.58$-$0.62$ in the statistical steady state, depending on the side-boundary condition. In our experiment, the measured mean spacings are $\lambda'=0.45\pm0.06$ in stage 1, $\lambda'=0.39\pm0.07$ in stage 2, and $\lambda'=0.30\pm0.02$ in the quasi-steady stage 3. Applying the no-slip to free-slip correction discussed in our scaling argument, which doubles the equivalent spacing, gives $\lambda'\approx0.61$ for our stage-3 quasi-steady state. Despite the differences in heating mode, viscosity law, and boundary conditions, this corrected quasi-steady spacing is consistent with the range inferred from \citeA{parmentier2000three}.



\subsection{Implications for the Earth}
\label{sec:discuss_earth}
Having validated our scalings and inferences against prior experiments and theoretical expectations we now compare to geochemical and geophysical observations.

\subsubsection{Rising speed of mantle plumes}
\label{sec:discuss_speed}
Directly using the diffusive time scale $\tau_\mathrm{diff}$ to scale the experimentally determined plume travel time $\tau_\mathrm{plume}$ to the Earth, we find that a $\tau_\mathrm{plume}=500$ s corresponds to 200 Myr, yielding a rising speed in the mantle $U_\mathrm{plume}^{m}$ of 1.5 cm/yr. This estimate is on the lower end of expectations both when comparing to theoretical, experimental, and observational expectations \cite<e.g.,>{Campbell2005}. A $U_\mathrm{plume}^{m}$ of 1.5 cm/yr is barely faster than some of the slowest spreading rates, and slower than Stokes terminal velocity from prior experimental and theoretical determinations ($\sim5$ cm/yr) \citeA{happel2012low,olson1985creeping,Richards1988}. It would also imply a lower mantle significant more viscous ($\sim10^{23}$ Pa$\cdot$s) than geophysical inversions. A diffusive time scale may not be the relevant characteristic time scale for thermals. 

A more appropriate scaling of $U_\mathrm{plume}$ should be based on a metric more relevant to the plume itself, for example, the P\'eclet number. We can define a P\'eclet number $Pe$ following \citeA{Cagney2015} for the plume as the ratio between diffusive and advection timescales:
\begin{equation}
  Pe = \frac{\tau_{\mathrm{diff}_\mathrm{plume}}}{\tau_{\mathrm{advect}_\mathrm{plume}}} = \frac{D_\mathrm{head}^2/\kappa}{D_\mathrm{head}/U_\mathrm{plume}} = U_\mathrm{plume}D_\mathrm{head}/\kappa
  \label{eqn:Pe}
\end{equation}
where $D_\mathrm{head}$ is the diameter of the fully developed plume head (right before hitting the top surface). With $U_\mathrm{plume}=1$ mm/s and $D_\mathrm{head}=110$ mm, we find $Pe=1048$ in our experiment (and $Pe=524$ for $U_\mathrm{plume}=0.5$ mm/s). 

Assuming the effective $Ra$ is the same between our experiment and the mantle, we can interpret this $Pe$ directly. For a mantle plume with a head diameter of 1000 km \cite{Campbell2005}, the corresponding fastest $U_\mathrm{plume}^{m}=$ 3.3 cm/yr, more than double our previous estimate. A diameter of a 1000 km is a reasonable value consistent with the maximum scaled diameter of our plume heads ($\sim$ 1200 km). The plume travel time would now be less than 90 Myr, compatible with the estimates in \citeA{Campbell2005} (100 Myr). It is also within the range found by \citeA{Bourdon2006} of 2 to 6 cm/yr based on geochemical modeling, for low buoyancy flux mantle plumes. 

This scaled rising speed can be much larger if the effective $Ra$ in the mantle is higher. For example, \citeA{Cagney2015} suggests that:

\begin{equation}
  Pe=3.82\times10^{-5}Ra+24.3
  \label{eqn:Pe_Ra}
\end{equation}
using the results of their single plume experiment conducted under similar conditions to our study, except for the spot heating. They calculate a rising speed of 1.28 cm/yr when scaled to a mantle of $Ra=10^7$. 

Given there is enough similarity between the spot-heated \cite{Cagney2015} and basally heated plumes in our experiment (section \ref{sec:compare_single}), we can modify equation \ref{eqn:Pe_Ra} by a prefactor that accounts for the larger speeds ($U_\mathrm{plume}$) and heating supply area ($D_{head}$) in the basally heated experiments compared to the spot-heating single plume case . In \citeA{Cagney2015}, $U_\mathrm{plume}\sim$0.12 mm/s and $D_\mathrm{head}=68$ mm, while our $U_\mathrm{plume}$ ranges from 0.5 to 1 mm/s, as much as an order magnitude more and $D_\mathrm{head}=110$ mm is nearly double. Based on equation \ref{eqn:Pe}, our $Pe$ should be 6.4 to 12.8 of that in \citeA{Cagney2015}. This 6.4 to 12.8 scaling factor can be then applied to equation \ref{eqn:Pe_Ra}, to scale the 1.28 cm/yr plume rising speed from \citeA{Cagney2015}. With $Ra=10^7$, the expected mantle plume speed will now be 8.2 cm/yr to 16.4 cm/yr, and plume travel times will range from 37 Myr to 18 Myr. This faster speed aligns with expectations given the temperature dependence of mantle viscosity and is consistent with dynamical estimates \cite<e.g.,>{Steinberger2006,Arnould2020}. If we further consider the change of the plume head size near the surface later in the experiment ($\sim$~65 mm instead of 110 mm), the lower end of the plume speed estimate will drop another 40\%, i.e., 4.9 cm/yr, and 62 Myr.

The large range in plume rising speeds and travel time estimates obtained by using scaling based on $\tau_\mathrm{diff}$ as the characteristic timescale or $Pe$ suggests caution in selecting the proper characteristic length and timescales during scaling. For the Earth's mantle, as convection dominates the heat transfer instead of diffusion, the $Pe$ is a better choice for capturing the faster plume speed.

For subsequent sections \ref{sec:discuss_dynamics} we use relative, rather than absolute timescales. We define the plume rising time $\tau_\mathrm{plume}$ as the characteristic timescale. Then, for the diffusion stage ($\tau_\mathrm{TBL}$), the result is 1$\tau_\mathrm{plume}$, 2$\tau_\mathrm{plume}$ represents the time for a plume cycle, and $\sim$  0.5$\tau_\mathrm{plume}$ for inter-plume interaction. The time needed for a full plume merging event is up to 1.34$\tau_\mathrm{plume}$ (Fig. \ref{fig:full_merge}).

\subsubsection{Width of mantle plumes}
In our experiment, the width of the plume stem (as captured by the material interface) is consistently $\sim$ 6 mm in the absence of inter-plume interactions. Scaled to the mantle it yields a 65 km stem diameter, about 30 to 60$\%$ of previous estimates \cite<e.g.,>{griffiths1991dynamics}. If the thermal halo of the plume is three times as wide as the plume width determined by the material interface \cite<e.g.,>{Newsome2011}, the width of the thermal anomaly is $\sim 200$ km,  challenging to resolve in either global or regional seismic tomographic imaging of the lower mantle \cite{French2015,Maguire2018}. This result underscores the importance of the methodology we have constructed. Without it, it would have not been possible to define the width of the plume stem to be so narrow. Numerical simulations analyzed using the temperature field alone (without Lagrangian analysis, \cite<e.g.,>{zhong2005dynamics,Arnould2020} could not capture it even at higher resolutions that computationally possible today. While it is hard to directly verify the existence of such purely thermal lower mantle plumes, the dynamics measured in the laboratory are still useful to understand the differences with upwellings resolved in seismic tomography \cite{French2015}, which may correspond to thermochemical plumes.

\subsubsection{Spacing of mantle plumes}

We find the minimum plume spacing to be 38 mm in stage 3, equivalent to $\sim$400 km in the mantle. Under free-slip boundary conditions, more representative of the core-mantle boundary, the equivalent spacing doubles to $\sim$800 km (cf. Fig.~8 in \citeA{Lister1989}). This matches the 800-1000 km minimum spacing of most seismic velocity minima at the core--mantle boundary, both within and outside the two Large Low Shear Velocity Provinces (LLSVPs, \citeA{Davaille2020}) interpreted as hot plumes. Plume-like features with similar spacing appear in both global \cite<e.g.,>{French2015,Mousavi2021} and regional \cite<e.g.,>{Suzuki2020} tomographic models, suggesting that plume clusters may be favored over large, coherent piles as explanations for LLSVPs.

\subsubsection{Morphological Evolution and Relation to Overall Flow}
\label{sec:discuss_dynamics}
\textbf{Merging and Branching}. Plumes rising and merging from the bottom (not near the top as in \citeA{Moses1993}) is hard to resolve clearly prior to our experiment \cite<e.g.,>{Olson1988,jellinek2003plume} but have been recognized in numerical simulations of global mantle convection which include radial and lateral viscosity variations \cite<e.g.,>{Arnould2020}. Under such conditions, a rising plume spends most of its time in the lower mantle. In \citeA{Arnould2020}, 40 to 50 Myr are needed for plumes to rise through the mantle (see their Fig. S6), and 60 to 70 Myr for two plumes to merge entirely from the bottom to the top (see their Fig. 7). Full merging times are $>1.4$ times the plume rising time, consistent with our results (1.34$\tau_\mathrm{plume}$). This can have significant implications for future geochemical mapping because of the consequences for material entrainment, as well as interpretations of seismic images.

The flow within the bottom-up merging plumes is dominantly vertical, so horizontal material transfer between them is limited. Consequently, the geochemical signals carried by the two plume heads should reflect distinct portions of the TBL, corresponding to the roots of the two plumes. This, however, requires both plume heads to reach the surface and to generate separate hotspot tracks in their life cycle.

The plate motion will further complicate the surface manifestation of merging plumes. For example, if the plate motion is perpendicular to the bottom-up style merging direction, the corresponding hotspot tracks could start with a distance larger or comparable to $\lambda_\mathrm{min}$, i.e., 800-1000 km, then gradually merge, with a timescale up to the plume rising time. It is however not easy to robustly recognize any sign of plume merging from the tracks, or from geophysical imaging \cite{lu2024synoptic}.


Plume merging tends to happen within a certain critical distance \citeA{Schubert2004}, e.g., $\lambda_\mathrm{min}$. However, wide-angle branching controlled by downwellings (Fig. \ref{fig:plume_dynamics_cartoon}), may help maintain the distance between plume heads, with a life time comparable to the plume travel time or more (cf. \ref{sec:branch}). This could be a possible mechanism for the long-lived parallel hotspots tracks with distinct geochemical signals ($>70$ Myr, \citeA{Buff2021}). 

\textbf{Detachment}. The observed detachment of the plume head (Fig. \ref{fig:plume_dynamics_cartoon}), sometimes referred to as a ``dying plume" \cite{Silveira2006} confirms the experimental observations in \citeA{Davaille2005}. It emphasizes the difference between transient, often episodic, plume dynamics and the present-day snapshot of Earth's interior that seismic tomography provides. An active hotspot with focused slow seismic velocity beneath it that does not extend all the way to the CMB, does not rule out a deep origin at the CMB.

Detachment is associated with plume pulsing, that is temporal changes in buoyancy flux. As mentioned in section \ref{sec:timescales}, the timescale of such pulsing activity (Fig. \ref{fig:Uy_detach}) is $\sim\tau_\mathrm{plume}/2$, comparable to the plume interaction timescale (Fig. \ref{fig:velocity_center}). We expect the pulsing to be ubiquitous and originate from fluctuations in the TBL thickness, after merging or splitting of plumes drains the buoyancy supply \cite{scott1986observations,loper1986mantle,olson1986solitary,Schubert1989}. Without invoking additional mechanisms, pulsing can explain $20\sim30$ Myr variations in volcanic activity at hotspots and nearby mid-ocean ridges \cite<e.g.,>{Vogt1974,Vogt1979}, if $\tau_\mathrm{plume} \sim 50$ Myr (cf. section \ref{sec:discuss_speed}, and \citeA{Arnould2020}). 

The shifting and motion of plumes may be promoted by the free-slip boundary condition at CMB to be as fast as the plume rising speed \cite{Schubert2004}, especially in the presence of subducting slabs \cite{Arnould2020}. Hence the timescale of disturbance in the TBL and associated pulsing could be shorter, and explain observed variations of a few Myr \cite<e.g.,>{Parnell2014}.

\textbf{Splitting}.  As during merging, splitting can contribute to fluctuations in the TBL. Disturbances in the boundary layer itself can also trigger splitting (Fig. \ref{fig:plume_dynamics_cartoon}). The splitting of plumes from the bottom TBL, before any interaction with structures near the top TBL \cite<e.g., cratons, >{Koptev2016}; or splitting from rheological structure \cite{Liu2020}, has not been seen experimentally to the best of our knowledge.

Splitting, much like branching could help maintain the distance between two neighboring plumes. This allows the plumes to stay apart at a critical distance comparable and below $\lambda_\mathrm{min}$. In fact, a continuous spectrum of inter-plume distance from zero to more than $\lambda_\mathrm{min}$ can be described as the consequence of fully merging, branching, splitting and isolated plumes.

Instead of a one-way road of irreversible merging, the splitting now provides a new evolutionary pathway and forms a cycle with merging (Fig. \ref{fig:plume_dynamics_cartoon}), redistributing the buoyancy supply from the TBL. 

In this respect the twin hotspots of Arago (also known as Rurutu) and Macdonald located in the hotspot highway \cite{Buff2021} might warrant further investigation. Lavas at these volcanoes sample the same geochemical source (HIMU) but are and have been only $\sim 1200$ km apart at the surface (corresponding to $<$ 600 km at the core-mantle boundary) for the last 70 Myr. 

Overall, the dynamic splitting, merging, and branching of plumes complicate the interpretation of plume-like seismic anomalies as distinct, connected, merging, or separating plumes.

\textbf{Superplume}. Our simple Rayleigh B\'enard experiment in a fluid with temperature-dependent viscosity and no other complexities, still produces not only merging, branching and splitting but also the ultimate superplume phenomena seen in numerical simulations \cite<e.g.,>{Arnould2020,Liu2020}. 

The tank walls (and temperature dependent viscosity, \citeA{zhang1997non}) promote the large-scale flow responsible for branching and superplume aggregation, given the constrained volume. At first glance it might not seem comparable to Earth's mantle, but in fact the walls can be regarded as an analog to the subduction curtain. These are areas of the mantle shielded by subducted slabs at the ocean basin scale (e.g. Pacific, \citeA{hayes2018slab2}). 

Our results suggest that the temperature-dependent viscous flow of Rayleigh-B\'enard convection can already explain much of the interconnected plumes and plume tree structure near LLSVPs imaged by seismic tomography \cite<e.g.,>{French2015,Tsekhmistrenko2021}. The relative absence of slow seismic velocities outside the LLSVPs,  as shown in Fig. \ref{fig:super_plume}a, might be due to the large-scale flow pushing plumes together (e.g.,\citeA{Androvandi2011}) to form clusters and composite superplumes. The wide ranges in plume spacing determined in our experiment may further explain variations in the spacing of slow seismic velocities in the lower mantle, particularly those outside the LLSVPs \cite{Suzuki2021} and far apart from each other ($>$2000 km separation distance  projected to the surface).



\subsubsection{Other timescales}
The current experiment could not robustly resolve timescales above 1000 s (cf. section \ref{sec:timescales}) given the temporal resolution of the SPIV system. Nevertheless, an attempt can be made, to look at the transition from more quiescent plume activities to burst of plumes, e.g. stage 1 plus stage 2. This timescale is $\sim3650$ s, or 7.3$\tau_\mathrm{plume}$. Taking $\tau_\mathrm{plume}$ as 40 to 50 Myr, such transition can be as long as $\sim300$ Myr. Together with the episodic plume activity (Fig. \ref{fig:number}), it may have implications on the influence of plumes on the supercontinent cycle \cite<e.g.,>{Doucet2020} or plate speeds \cite{Lithgow-Bertelloni1998}. Future longer experiments could probably reveal more stages of transient or quasi-stable plume behavior and spatial self-reorganization.

The visual inspection of cluster evolution revealed a survival or persistence timescale for individually isolated plumes of $\sim1200$ s, which when scaled using $Pe$ (cf. \ref{eqn:Pe}) suggests a survival time of 50-100 Myr, nearly twice or more of the travel (or rise) time. This estimate is consistent with long-lived hotspots whose first eruptive products date back to $\sim100$ Myr (personal communication with Matthew G Jackson) such as Hawaii (80 Myr, \citeA{Tarduno2003ScienceEmperor}), Samoa (110 Myr, \citeA{Jackson2024AGUAdvSamoaOJP}), Louisville (120 Myr, \citeA{konter2025pacific}), and Tristan-Gough (140 Myr, \citeA{Rohde2013TectoTGW}).



\subsection{Limitations and future improvements}
Despite of the extensive velocity data coverage in space and time in our experiment, the quality of the plume analysis is still limited, especially by the temporal resolution and near-boundary flow \cite{bao2025piv}. We did not measure the temperature within the fluid, an achievable future direction given our experimental and postprocessing capabilities \cite{Cagney2015}. We also see similar plume behaviors in other experiments performed in the same system, with detailed configuration and statistics a subject of future work \cite{Bao2024PhD}. Besides performing a different set of experiments, it is also possible to release the full potential of the current 4D velocity measurements, using computational approaches such as automatic time-dependent adjoint methods \cite{mitusch2019dolfin} which can help us obtain the temperature field and perform Lagrangian material tracking at high resolution to study entrainment and perform surface-to-source mapping of anomalies in hotspot lavas. We will explore this challenging, yet exciting direction in an upcoming paper. 

\section{Conclusion}
Powered by 4D velocity data and a methodology tailored toward fully characterizing plume interaction and evolutionary dynamics, we are able to systematically evaluate the velocity and material transport of viscous plumes and their spatiotemporal distribution in unprecedented detail. We show that with temperature-dependent viscosity alone, the plumes in Rayleigh B\'enard convection already exhibit rich dynamical behavior and varied morphology. As a result, we are able to propose a network of possible evolutionary pathways, which can impact our view of material entrainment and surface geochemical anomalies and the interpretation of recent seismic imaging. 

Our results confirm and expand on previous results including theoretical and experimental estimates of plume lengthscales and timescales.  We use these scalings to explain multiple mantle-plume related observations from the configuration of hotspot tracks, to the 3D interconnected tree-like low seismic velocity anomaly structure under Africa, and the spacing of seismic velocity local minima in the lowermost mantle. Taken together, it is therefore possible, using the plume cluster results, to explain many aspects of the slow seismic velocity in the lower mantle inside and outside LLSVPs. This work also opens the door to more quantitative comparison and benchmarking  between geodynamical laboratory experiments and numerical simulations. 

The rich spatio-temporal variations in plume morphology and motions we observed confirms previous conclusions on the inevitability of plume motion \cite{Olson1988,Gonnermann2004,Arnould2020}. It highlights how unlikely long-lived static plumes as imagined by \citeA{Morgan1971} might be in the convecting mantle. They are thin and vulnerable to disruption from the large-scale flow. At any given time individual plumes that are not merging, splitting or clustering in a superplume, could vary from 0-60\% of the total number of plumes. Geochronological and geochemical data confirm that hotspots indeed move with respect to each other, sometimes significantly even within an ocean basin \cite{molnar1987relative,stock2003hotspots,oneill2003geodynamic,konrad2018relative}. However, it leaves unexplained how long-lived hotspots such as Hawaii can remain in place for over 100 Myr. It suggests that other anchoring mechanisms must exist, perhaps related to thermochemical convection \cite{Davaille2002,Jellinek2002,lebars2004whole}, rooting plumes in ULVZs \cite{williams1996seismic,lay1998core} or others not yet discussed.

Our experiments may be further improved with better spatiotemporal coverage and resolution, using either updated experimental systems or a digital twin constrained by the current experimental data, to achieve more fine-grained plume analysis. In the future, specially-designed experiments and simulations can explore the zoo of plume dynamical behaviors under more controlled conditions, and entrainment analysis and metrics developed to explore surface-to-source mapping of geochemical anomalies of hotspot lavas to their geographical location in the deep mantle.

\section*{Availability Statement}
Jupyter Notebooks for the plume analysis can be found at \citeA{bao2025z1}.

\section*{Conflict of Interest declaration}
The authors declare there are no conflicts of interest for this manuscript.

\acknowledgments
The project was directly supported by the National Science Foundation under grant EAR-1900633 to C.L-B. C.L-B. was further supported by the Louis B. and Martha B. Slichter Endowed Chair in Geosciences. X.B. was also supported by Harvard Reginald A. Daly Postdoctoral Fellowship.

\appendix

\section{Thermal Boundary Layer thickness and its change over time}
\label{app:TBL}

This appendix supports the discussion in section \ref{sec:discuss_thickness_tau_head}. We first clarify the definitions of the TBL initiation time $\tau_{TBL}$, the TBL/stem thickness $\delta$, the prefactor $p$, and the nondimensional temperature level $\Theta$ used to identify an isotherm-defined TBL edge. We then apply these definitions to the present experiment and to the \citeA{Davaille2005} experiment. Finally, we examine whether the lower TBL thickness is expected to change as the fluid interior becomes progressively warmer.

\subsection{The definition of $\tau_{TBL}$ and $\delta$, and comparison with Davaille and Vatteville (2005)}

The classical TBL analysis in \citeA{Howard1966} is often written as
\begin{equation}
\tau_{TBL}
=
\frac{H^2}{p\kappa}
\left(
\frac{Ra_c}{Ra}
\right)^{2/3},
\label{eq:tau_tbl_general}
\end{equation}
or equivalently
\begin{equation}
\delta
=
\sqrt{p\kappa \tau_{TBL}}
=
H
\left(
\frac{Ra_c}{Ra}
\right)^{1/3}.
\label{eq:delta_general}
\end{equation}
Here $p$ is not a universal stability constant. The value $p=\pi$ comes from defining the transient conductive boundary-layer thickness from the basal heat flux, or equivalently from the basal temperature gradient. For a semi-infinite layer subject to a sudden bottom-temperature jump,
\begin{equation}
\Theta(z,t)
\equiv
\frac{T(z,t)-T_i}{T_b-T_i}
=
\mathrm{erfc}
\left(
\frac{z}{2\sqrt{\kappa t}}
\right),
\label{eq:erfc_solution}
\end{equation}
where
\begin{equation}
\frac{d}{d\eta}\mathrm{erfc}(\eta)
=
-\frac{2}{\sqrt{\pi}}\exp(-\eta^2).
\end{equation}
Evaluating the gradient of equation \eqref{eq:erfc_solution} at the bottom boundary gives
\begin{equation}
\left.
\frac{\partial T}{\partial z}
\right|_{z=0}
=
-\frac{T_b-T_i}{\sqrt{\pi\kappa t}},
\end{equation}
and the conductive heat flux is
\begin{equation}
q(t)
=
k\frac{T_b-T_i}{\sqrt{\pi\kappa t}}.
\end{equation}
Now define $\delta_q$ as the thickness of a hypothetical linear profile that has the same basal temperature contrast and the same basal gradient as the actual $\mathrm{erfc}$ solution. Equivalently, $\delta_q$ is the distance over which the tangent to the $\mathrm{erfc}$ profile at $z=0$ must be extrapolated before reaching the far-field temperature $T_i$. This linear profile has flux
\begin{equation}
q(t)=k\frac{T_b-T_i}{\delta_q}.
\end{equation}
Equating the two fluxes gives
\begin{equation}
\delta_q=\sqrt{\pi\kappa t}.
\end{equation}
Thus, if the conductive thickness is written as $\delta_q=\sqrt{p\kappa t}$, this particular flux-equivalent, or basal-slope-equivalent, definition gives $p=\pi$. The value $p=\pi$ therefore does not describe a true geometric thickness of the thermal anomaly. A true geometric thickness requires choosing a particular isotherm, for example $\Theta=\Theta^\ast$, whose position is
\begin{equation}
z
=
2\sqrt{\kappa t}\,
\mathrm{erfc}^{-1}(\Theta^\ast),
\end{equation}
and would lead to a different numerical prefactor. Hence $p=\pi$ should not be interpreted as the thickness of a visible isotherm or a material interface.

For an isotherm-defined thickness, equation \eqref{eq:erfc_solution} gives
\begin{equation}
    \delta_\Theta
    =
    2\sqrt{\kappa t}\,
    \mathrm{erfc}^{-1}(\Theta),
    \label{eq:delta_isotherm}
\end{equation}
or, if written as $\delta_\Theta=\sqrt{p_\Theta \kappa t}$,
\begin{equation}
    p_\Theta
    =
    4
    \left[
        \mathrm{erfc}^{-1}(\Theta)
    \right]^2.
    \label{eq:p_isotherm}
\end{equation}
Therefore the inferred value of $p$ depends on which temperature level is used to define the TBL front.

For the present experiment, we use
\begin{equation}
    T_i=25^\circ\mathrm{C},
    \qquad
    T_b=80^\circ\mathrm{C},
    \qquad
    \Delta T = 55^\circ\mathrm{C},
\end{equation}
\begin{equation}
    H=0.275~\mathrm{m},
    \qquad
    \kappa=1.05\times 10^{-7}~\mathrm{m^2\,s^{-1}}.
\end{equation}
The cold/global Rayleigh number and hot-bottom/local Rayleigh number are
\begin{equation}
    Ra_{\mathrm{global}}=Ra=1.9\times 10^6,
    \qquad
    Ra_{\mathrm{local}}=Ra_\delta=1.2\times 10^8.
\end{equation}
Unless otherwise stated, the reference critical Rayleigh number is
\begin{equation}
    Ra_c=1708.
\end{equation}

\subsubsection{Plume-initiation timescale: 500 s versus ramp-corrected time}

From the PIV/PSD analysis, the first plume-initiation timescale is about
\begin{equation}
    \tau_{TBL}^{\mathrm{obs}}\simeq 500~\mathrm{s}.
\end{equation}
However, the bottom temperature did not jump instantaneously to $80^\circ\mathrm{C}$; it increased approximately linearly from $25^\circ\mathrm{C}$ to $80^\circ\mathrm{C}$ over
\begin{equation}
    t_r = 250~\mathrm{s}.
\end{equation}
A simple midpoint correction therefore gives the effective growth time
\begin{equation}
    \tau_{\mathrm{eff}}
    \simeq
    500-\frac{t_r}{2}
    =
    375~\mathrm{s}.
    \label{eq:half_ramp_time}
\end{equation}

A more explicit diffusion calculation gives a similar effective sudden-jump time. For a linear ramp followed by a constant temperature, the semi-infinite conductive solution at $t>t_r$ is
\begin{equation}
    \Theta_{\mathrm{ramp}}(z,t)
    =
    \frac{1}{t_r}
    \int_{0}^{t_r}
    \mathrm{erfc}
    \left[
        \frac{z}{2\sqrt{\kappa(t-s)}}
    \right] ds
    =
    \frac{1}{t_r}
    \int_{t-t_r}^{t}
    \mathrm{erfc}
    \left[
        \frac{z}{2\sqrt{\kappa u}}
    \right] du .
    \label{eq:ramp_solution}
\end{equation}
We can define an equivalent sudden-jump time $t_{\mathrm{eq}}(z)$ by
\begin{equation}
    \mathrm{erfc}
    \left[
        \frac{z}{2\sqrt{\kappa t_{\mathrm{eq}}}}
    \right]
    =
    \Theta_{\mathrm{ramp}}(z,500~\mathrm{s}).
    \label{eq:teq_definition}
\end{equation}
For the dynamically relevant depths,
\begin{equation}
    t_{\mathrm{eq}}(z=6.66~\mathrm{mm}) \simeq 366~\mathrm{s},
\end{equation}
\begin{equation}
    t_{\mathrm{eq}}(z=7.5~\mathrm{mm}) \simeq 367~\mathrm{s}.
\end{equation}
For a much colder outer penetration front corresponding to the global-Rayleigh-number thickness, $z\simeq 26.5$ mm, the same calculation gives
\begin{equation}
    t_{\mathrm{eq}}(z=26.5~\mathrm{mm}) \simeq 394~\mathrm{s}.
\end{equation}
Thus the simple half-ramp correction, $\tau_{\mathrm{eff}}\simeq 375$ s, is adequate to first order. Depending on which diagnostic depth is used, the exact ramp-to-step equivalent time is roughly $365$-$395$ s.

Equivalent heat-flux and heat-content definitions give comparable values. For the wall heat flux,
\begin{equation}
    q_{\mathrm{ramp}}(t)
    =
    \frac{k\Delta T}{t_r}
    \int_0^{t_r}
    \frac{ds}{\sqrt{\pi\kappa(t-s)}}
    =
    \frac{2k\Delta T}{t_r\sqrt{\pi\kappa}}
    \left(
        \sqrt{t}-\sqrt{t-t_r}
    \right),
\end{equation}
so that equating $q_{\mathrm{ramp}}(500~\mathrm{s})$ to the sudden-step flux gives
\begin{equation}
    t_{\mathrm{eq},q}
    =
    \left[
        \frac{t_r}
        {2\left(\sqrt{t}-\sqrt{t-t_r}\right)}
    \right]^2
    \simeq
    364~\mathrm{s}.
\end{equation}
For the integrated conductive heat content, the equivalent sudden-jump time is
\begin{equation}
    t_{\mathrm{eq},Q}
    =
    \left[
        \frac{2}{3t_r}
        \left(
            t^{3/2}-(t-t_r)^{3/2}
        \right)
    \right]^2
    \simeq
    371~\mathrm{s}.
\end{equation}

\subsubsection{Thickness estimates in the present experiment}

Using equation \eqref{eq:delta_general} with $Ra_c=1708$,
\begin{equation}
    \delta_{\mathrm{global}}
    =
    H
    \left(
        \frac{Ra_c}{Ra_{\mathrm{global}}}
    \right)^{1/3}
    =
    0.275
    \left(
        \frac{1708}{1.9\times10^6}
    \right)^{1/3}
    \simeq
    26.5~\mathrm{mm}.
\end{equation}
By contrast,
\begin{equation}
    \delta_{\mathrm{local}}
    =
    H
    \left(
        \frac{Ra_c}{Ra_{\mathrm{local}}}
    \right)^{1/3}
    =
    0.275
    \left(
        \frac{1708}{1.2\times10^8}
    \right)^{1/3}
    \simeq
    6.66~\mathrm{mm}.
\end{equation}

The two choices imply very different temperature levels if interpreted through the conductive profile. Using
\begin{equation}
    \Theta_\delta(t)
    =
    \mathrm{erfc}
    \left[
        \frac{\delta}{2\sqrt{\kappa t}}
    \right],
\end{equation}
we obtain:
\begin{center}
\begin{tabular}{lccccccc}
\hline
Ra choice &
$\delta$ &
$p$ at 375 s &
$p$ at 500 s &
$\Theta$ at 375 s &
$T$ at 375 s &
$\Theta$ at 500 s &
$T$ at 500 s \\
\hline
$Ra_{\mathrm{global}}$ &
26.5 mm &
17.9 &
13.4 &
0.0028 &
$25.15^\circ$C &
0.0096 &
$25.53^\circ$C \\
$Ra_{\mathrm{local}}$ &
6.66 mm &
1.13 &
0.85 &
0.453 &
$49.9^\circ$C &
0.515 &
$53.3^\circ$C \\
\hline
\end{tabular}
\end{center}

Thus the global-Rayleigh-number thickness corresponds to an extremely cold, far outer conductive penetration front, whereas the local-Rayleigh-number thickness corresponds to a mid-temperature, dynamically active hot plume-root layer. This distinction is important in strongly temperature-dependent viscosity experiments. The same global $Ra$ does not imply the same hot-bottom TBL thickness in isoviscous and strongly temperature-dependent viscosity fluids.

\subsubsection{Meaning of the observed 7.5 mm depth}

Our observed TBL proxy, the upper edge of a just initiated plume head, is at about
\begin{equation}
    z=7.5~\mathrm{mm}.
\end{equation}
From equation \eqref{eq:erfc_solution},
\begin{equation}
    \Theta(7.5~\mathrm{mm},375~\mathrm{s})
    =
    \mathrm{erfc}
    \left[
        \frac{0.0075}{2\sqrt{(1.05\times10^{-7})(375)}}
    \right]
    \simeq
    0.398,
\end{equation}
corresponding to
\begin{equation}
    T
    =
    25+0.398\times55
    \simeq
    46.9^\circ\mathrm{C}.
\end{equation}
Similarly,
\begin{equation}
    \Theta(7.5~\mathrm{mm},500~\mathrm{s})
    \simeq
    0.464,
\end{equation}
corresponding to
\begin{equation}
    T
    \simeq
    50.5^\circ\mathrm{C}.
\end{equation}
Therefore the observed 7.5 mm scale corresponds approximately to a $\Theta\simeq0.4$--0.46 isotherm, not to a cold outer thermal front.

For reference, at $t=375$ s in the present experiment, we have the following:
\begin{center}
\begin{tabular}{ccc}
\hline
$\Theta$ & $T$ & $z$ from bottom \\
\hline
0.1 & $30.5^\circ$C & 14.6 mm \\
0.2 & $36.0^\circ$C & 11.4 mm \\
0.3 & $41.5^\circ$C & 9.20 mm \\
0.4 & $47.0^\circ$C & 7.47 mm \\
0.5 & $52.5^\circ$C & 5.99 mm \\
\hline
\end{tabular}
\end{center}

This table confirms that a $7.5$ mm TBL scale corresponds closely to the $\Theta\simeq0.4$ isotherm for the ramp-corrected growth time.

\subsubsection{What happens if $p=\pi$ is imposed?}

If $p=\pi$ and the local Rayleigh number is used,
\begin{equation}
    \tau_{TBL}
    =
    \frac{H^2}{\pi\kappa}
    \left(
        \frac{Ra_c}{Ra_{\mathrm{local}}}
    \right)^{2/3}
    \simeq
    135~\mathrm{s}
    \qquad
    (Ra_c=1708),
\end{equation}
which gives a reasonable hot-layer thickness but underpredicts the observed initiation time.

If $p=\pi$ and the global Rayleigh number is used,
\begin{equation}
    \tau_{TBL}
    =
    \frac{H^2}{\pi\kappa}
    \left(
        \frac{Ra_c}{Ra_{\mathrm{global}}}
    \right)^{2/3}
    \simeq
    2135~\mathrm{s},
\end{equation}
and
\begin{equation}
    \delta\simeq 26.5~\mathrm{mm}.
\end{equation}
This is much too thick and too slow for the observed plume geometry. Thus $p=\pi$ cannot simultaneously explain the observed initiation time and the observed dynamically active hot-layer thickness.

\subsubsection{Comparison to Davaille and Vatteville with newly supplied properties}

For the \citeA{Davaille2005} experiment, the published setup is a sugar syrup layer initially at
\begin{equation}
    T_i=21^\circ\mathrm{C}
\end{equation}
and suddenly heated from below at
\begin{equation}
    T_b=53^\circ\mathrm{C},
    \qquad
    \Delta T=32^\circ\mathrm{C},
\end{equation}
with the plume outlined by the $24.6^\circ$C isotherm and a reported Rayleigh number
\begin{equation}
    Ra\simeq 1.7\times10^6.
\end{equation}
The onset time is
\begin{equation}
    t_c\simeq 370~\mathrm{s},
\end{equation}
and the detachment/cycle time is
\begin{equation}
    t\simeq 550~\mathrm{s}.
\end{equation}

Using the newly supplied properties at $21^\circ$C,
\begin{equation}
    k = 0.376+5.25\times10^{-4}T
    \simeq
    0.3870~\mathrm{W\,m^{-1}\,K^{-1}},
\end{equation}
\begin{equation}
    C_p = 2345+40.7T
    \simeq
    3199.7~\mathrm{J\,kg^{-1}\,K^{-1}},
\end{equation}
\begin{equation}
    \rho=1388+5.74\times10^{-4}\times(20-T)\simeq 1388~\mathrm{kg\,m^{-3}},
\end{equation}
so that
\begin{equation}
    \kappa
    =
    \frac{k}{\rho C_p}
    \simeq
    8.71\times10^{-8}~\mathrm{m^2\,s^{-1}}.
\end{equation}
The viscosity law gives
\begin{equation}
    \eta(21^\circ\mathrm{C})
    \simeq
    5.35~\mathrm{Pa\,s},
\end{equation}
and
\begin{equation}
    \eta(53^\circ\mathrm{C})
    \simeq
    0.325~\mathrm{Pa\,s}.
\end{equation}
With $H=0.148$ m, the initial/ambient/global Rayleigh number is
\begin{equation}
    Ra_{21}
    =
    \frac{\rho g\alpha\Delta T H^3}{\eta(21^\circ\mathrm{C})\kappa}
    \simeq
    1.74\times10^6,
\end{equation}
which reproduces the published $Ra=1.7\times10^6$.

If instead the bottom viscosity at $53^\circ$C is used, the local/hot Rayleigh number would be
\begin{equation}
    Ra_{53}
    =
    Ra_{21}
    \frac{\eta(21^\circ\mathrm{C})}{\eta(53^\circ\mathrm{C})}
    \simeq
    2.87\times10^7.
\end{equation}
This larger value is not the Rayleigh number reported for the \citeA{Davaille2005} transient-onset experiment. The parameters are summarized in Table \ref{tab:davaille}.

\begin{table*}
\caption{Fluid properties of the TBL experiment in \citeA{Davaille2005,Davaille2008}, updated using the newly provided material properties. Properties used for the global Rayleigh number are evaluated at the initial temperature, $T=21^\circ\mathrm{C}$. The local Rayleigh number is calculated using the basal viscosity at $T=53^\circ\mathrm{C}$.}
\label{tab:davaille}
\centering
\begin{tabular}{lcr}
\hline\hline
Properties & Symbol & Value \\
\hline
Density (21 $^\circ$C) & $\rho$ & 1388 $\mathrm{kg\,m^{-3}}$ \\
Dynamic viscosity (21/53 $^\circ$C) & $\eta$ & 5.345 / 0.3246 Pa$\cdot$s \\
Thermal expansion coefficient & $\alpha$ & $5.74\times 10^{-4}$ $\mathrm{K}^{-1}$ \\
Thermal conductivity (21 $^\circ$C) & $k$ & 0.3870 W$\cdot$m$^{-1}$K$^{-1}$ \\
Specific heat capacity (21 $^\circ$C) & $C_P$ & 3199.7 J$\cdot$kg$^{-1}$K$^{-1}$ \\
Thermal diffusivity (21 $^\circ$C) & $\kappa$ & $8.71\times 10^{-8}$ m$^2$/s \\
Thickness of the fluid & $H$ & 0.148 m \\
Thickness of the bottom TBL & $\delta$ & 0.0147 m \\
Global Rayleigh number, using $\eta(21^\circ\mathrm{C})$ & $Ra$ & $1.74 \times 10^6$ \\
Local Rayleigh number, using $\eta(53^\circ\mathrm{C})$ & $Ra_{\delta}$ & $2.87 \times 10^7$ \\
Time of plume initiation & $\tau_{TBL}$ & 370 s \\
\hline\hline
\end{tabular}
\end{table*}

For $Ra_c=1708$, the global and local choices give
\begin{equation}
    \delta_{\mathrm{global}}
    =
    H
    \left(
        \frac{1708}{1.74\times10^6}
    \right)^{1/3}
    \simeq
    14.7~\mathrm{mm},
\end{equation}
and
\begin{equation}
    \delta_{\mathrm{local}}
    =
    H
    \left(
        \frac{1708}{2.87\times10^7}
    \right)^{1/3}
    \simeq
    5.8~\mathrm{mm}.
\end{equation}
The corresponding effective $p=\delta^2/(\kappa t_c)$ and isotherm levels are
\begin{center}
\begin{tabular}{lcccc}
\hline
Definition & $\delta$ & $p$ at $t_c=370$ s & $\Theta$ & $T$ \\
\hline
Global $Ra_{21}$, $Ra_c=1708$ &
14.7 mm &
6.71 &
0.067 &
$23.1^\circ$C \\
Local $Ra_{53}$, $Ra_c=1708$ &
5.8 mm &
1.04 &
0.472 &
$36.1^\circ$C \\
Howard heat-flux thickness, $p=\pi$ &
10.1 mm &
$\pi$ &
0.210 &
$27.7^\circ$C \\
Observed/plotted $24.6^\circ$C isotherm &
12.7 mm &
5.04 &
0.1125 &
$24.6^\circ$C \\
\hline
\end{tabular}
\end{center}

Thus the $24.6^\circ$C isotherm in the \citeA{Davaille2005} figure corresponds to
\begin{equation}
    \Theta
    =
    \frac{24.6-21}{53-21}
    =
    0.1125,
\end{equation}
and therefore
\begin{equation}
    p_\Theta
    =
    4
    \left[
        \mathrm{erfc}^{-1}(0.1125)
    \right]^2
    \simeq
    5.04,
\end{equation}
\begin{equation}
    \delta_{24.6^\circ\mathrm{C}}
    =
    \sqrt{
        5.04\,
        (8.71\times10^{-8})
        (370)
    }
    \simeq
    12.7~\mathrm{mm}.
\end{equation}
If a $24.0^\circ$C isotherm is used instead,
\begin{equation}
    \Theta
    =
    \frac{24.0-21}{32}
    =
    0.09375,
\end{equation}
\begin{equation}
    p_\Theta\simeq 5.62,
    \qquad
    \delta_{24.0^\circ\mathrm{C}}
    \simeq
    13.5~\mathrm{mm}.
\end{equation}

This explains why different ``TBL thicknesses'' appear depending on the definition. The \citeA{Davaille2005} visible low-temperature front ($24$--$24.6^\circ$C) is a cold outer conductive isotherm ($\Theta\simeq0.1$), whereas the local-Rayleigh-number thickness corresponds to a much hotter level, $\Theta\simeq0.47$, close to the mid-temperature region of the hot feeding layer.

\subsubsection{Summary of operational TBL definitions}

The phrase ``TBL thickness'' is operational and definition-dependent. A heat-flux-equivalent thickness gives $p=\pi$. A visible isotherm gives $p=p_\Theta$, which depends on the chosen non-dimensional temperature $\Theta$. A critical-Rayleigh-number estimate gives $\delta=H(Ra_c/Ra)^{1/3}$, but its physical meaning depends on whether $Ra$ is defined using the cold/ambient viscosity or the hot/basal viscosity. In strongly temperature-dependent viscosity fluids, these choices are not interchangeable.

For the present experiment, using the hot-bottom/local Rayleigh number gives a thickness scale of $6$--$7$ mm, corresponding to $\Theta\simeq0.45$--0.52 for $t=375$--500 s. This is consistent with a dynamically active hot plume-root layer. Using the cold/global Rayleigh number gives a much thicker scale, about $26.5$ mm, corresponding only to a very cold outer penetration front, $\Theta\simeq0.003-0.01$. The observed $7.5$ mm depth corresponds to $\Theta\simeq0.40$--0.46, or $T\simeq47$--$51^\circ$C, and is therefore better interpreted as a hot-layer/plume-root diagnostic than as the full cold conductive penetration depth. One would, however, infer a thickness of 14.6 mm if $\Theta\simeq0.1$ is used as the TBL cutoff, thicker than that in \citeA{Davaille2005} of $\sim12.7$ mm.

For the \citeA{Davaille2005} experiment, the reported $Ra=1.7\times10^6$ is reproduced by using the initial/ambient $21^\circ$C viscosity. The visible $24.6^\circ$C isotherm corresponds to $\Theta=0.1125$, $p\simeq5.0$, and $\delta\simeq12.7$ mm. The hot-bottom/local Rayleigh-number thickness would instead be about $5.8$ mm and correspond to $\Theta\simeq0.47$, which is a different, much hotter, operational definition of TBL thickness, and is consistent with the $\Theta\simeq0.45$ derived from the local Rayleigh number in the present experiment. The corresponding prefactor $p=1.13$ from the local Rayleigh number in the present experiment is also consistent with $p=1.04$ in \citeA{Davaille2005}. Therefore, the local Rayleigh number reconciles the active TBL scale in the two experiments.

\subsection{TBL thickness change with progressively hotter fluid interior}

We next consider whether a progressively hotter fluid interior necessarily implies a thinner lower TBL. This is a separate question from the operational definition of $\delta$ above: here the definition of the active lower TBL is fixed, and we examine how the expected relative thickness changes as the bulk/interior temperature increases. We define a local TBL Rayleigh number as
\begin{equation}
    Ra_{\mathrm{TBL}}
    =
    \frac{\rho g \alpha \Delta T_{\mathrm{TBL}} \delta^3}
    {\kappa \eta_{\mathrm{eff}}},
    \label{eq:Ra_TBL_definition}
\end{equation}
where $\Delta T_{\mathrm{TBL}}$ is the temperature drop across the lower TBL, $\delta$ is the TBL thickness, and $\eta_{\mathrm{eff}}$ is an effective viscosity sampled by the active lower TBL. If TBL instability occurs at an approximately fixed effective boundary-layer Rayleigh number, then
\begin{equation}
    \delta
    =
    \left(
        \frac{Ra_{\mathrm{TBL}}\kappa\eta_{\mathrm{eff}}}
        {\rho g\alpha \Delta T_{\mathrm{TBL}}}
    \right)^{1/3}.
\end{equation}
For fixed $Ra_{\mathrm{TBL}}$,
\begin{equation}
    \delta
    \propto
    \left(
        \frac{\eta_{\mathrm{eff}}}
        {\Delta T_{\mathrm{TBL}}}
    \right)^{1/3}.
    \label{eq:delta_eta_over_DT}
\end{equation}
Thus the sign of the TBL-thickness change is controlled by the ratio $\eta_{\mathrm{eff}}/\Delta T_{\mathrm{TBL}}$, not by $\Delta T_{\mathrm{TBL}}$ alone.

We use the measured viscosity law for the present syrup,
\begin{equation}
    \eta(T)
    =
    \exp
    \left(
        4.642\times10^{-4}T^2
        -1.246\times10^{-1}T
        +6.325
    \right),
    \label{eq:viscosity_law_note}
\end{equation}
where $T$ is in $^\circ\mathrm{C}$ and $\eta$ is in Pa s. The bottom temperature is fixed at
\begin{equation}
    T_b=80^\circ\mathrm{C}.
\end{equation}
We define the bottom TBL edge using
\begin{equation}
    \Theta
    =
    \frac{T-T_i}{T_b-T_i},
    \label{eq:theta_bulk_reference}
\end{equation}
where $T_i$ is the evolving bulk/interior temperature. This definition differs slightly from the half-spacing cooling calculation elsewhere in this appendix, where the cold reference temperature is the top-plate temperature rather than the evolving interior temperature.

For the calculation below, the TBL edge is taken to be
\begin{equation}
    \Theta_e=0.1.
\end{equation}
Then
\begin{equation}
    T_e
    =
    T_i+\Theta_e(T_b-T_i),
\end{equation}
\begin{equation}
    \Delta T_{\mathrm{TBL}}
    =
    T_b-T_e
    =
    (1-\Theta_e)(T_b-T_i),
\end{equation}
and we estimate
\begin{equation}
    \eta_{\mathrm{eff}}
    \simeq
    \eta(T_{\mathrm{avg}}),
    \qquad
    T_{\mathrm{avg}}
    =
    \frac{T_b+T_e}{2}.
\end{equation}
The reference state is the initial interior temperature $T_i=25^\circ\mathrm{C}$. Relative thickness is computed from equation \eqref{eq:delta_eta_over_DT} as
\begin{equation}
    \delta_{\mathrm{rel}}
    =
    \frac{\delta}{\delta_{\mathrm{ref}}}
    =
    \left[
        \frac{
            \eta_{\mathrm{eff}}/\Delta T_{\mathrm{TBL}}
        }{
            \eta_{\mathrm{eff,ref}}/\Delta T_{\mathrm{TBL,ref}}
        }
    \right]^{1/3}.
    \label{eq:relative_delta_note}
\end{equation}
The relative conductive heat flux across the TBL is defined as
\begin{equation}
    q_{\mathrm{rel}}
    =
    \frac{q_{\mathrm{TBL}}}{q_{\mathrm{TBL,ref}}}
    =
    \frac{
        \Delta T_{\mathrm{TBL}}/\delta
    }{
        \Delta T_{\mathrm{TBL,ref}}/\delta_{\mathrm{ref}}
    }.
    \label{eq:relative_heat_flux_note}
\end{equation}

\begin{table}
\caption{Estimated change in lower-TBL thickness and heat flux as the interior temperature $T_i$ increases. The TBL edge is defined by $\Theta_e=0.1$, and $\eta_{\mathrm{eff}}$ is estimated from the endpoint-average temperature across the TBL.}
\label{tab:relative_delta}
\centering
\begin{tabular}{cccccccc}
\hline
$T_i$ ($^\circ\mathrm{C}$) &
$T_b-T_i$ &
$T_e$ &
$\Delta T_{\mathrm{TBL}}$ &
$T_{\mathrm{avg}}$ &
$\eta_{\mathrm{eff}}$ (Pa s) &
$\delta_{\mathrm{rel}}$ &
$q_{\mathrm{rel}}$ \\
\hline
25 & 55 & 30.5 & 49.5 & 55.25 & 2.36 & 1.00 & 1.00 \\
30 & 50 & 35.0 & 45.0 & 57.50 & 2.00 & 0.98 & 0.93 \\
35 & 45 & 39.5 & 40.5 & 59.75 & 1.71 & 0.96 & 0.85 \\
40 & 40 & 44.0 & 36.0 & 62.00 & 1.47 & 0.95 & 0.77 \\
45 & 35 & 48.5 & 31.5 & 64.25 & 1.27 & 0.94 & 0.67 \\
50 & 30 & 53.0 & 27.0 & 66.50 & 1.10 & 0.95 & 0.58 \\
55 & 25 & 57.5 & 22.5 & 68.75 & 0.954 & 0.96 & 0.47 \\
60 & 20 & 62.0 & 18.0 & 71.00 & 0.834 & 0.99 & 0.37 \\
65 & 15 & 66.5 & 13.5 & 73.25 & 0.733 & 1.04 & 0.26 \\
\hline
\end{tabular}
\end{table}

This calculation shows that, although $\Delta T_{\mathrm{TBL}}$ decreases substantially as the fluid interior warms, the effective viscosity sampled by the lower TBL also decreases because the whole TBL becomes warmer. These two effects largely compensate in the ratio $\eta_{\mathrm{eff}}/\Delta T_{\mathrm{TBL}}$. Under the fixed-$Ra_{\mathrm{TBL}}$ assumption, the inferred TBL thickness therefore remains nearly constant, varying by only a few percent over a wide range of interior temperatures. Using $\Theta_e=0.5$ instead of $\Theta_e=0.1$ gives the same conclusion.

The endpoint-average estimator of $\eta_{\mathrm{eff}}$ is not the only possible choice. \citeA{davaille1994onset} showed (their Equations (3), (5)--(6), and Figure~2) that for strongly temperature-dependent viscosity the instability of a growing boundary layer is controlled entirely by its low-viscosity side: the unstable sublayer spans temperatures within $\Delta T_{\mathrm{eff}} \simeq 2.24\,\Delta T_\nu$ of the mobile boundary, where $\Delta T_\nu = [-\partial\ln\eta/\partial T]^{-1}$ is the viscous temperature scale, and the boundary-layer scales are built on the viscosity of the mobile side rather than on an average across the whole layer. Their analysis is for a fluid cooled from above, where the mobile side is the interior; for our heated-from-below TBL the mobile side is the hot plate, so both scales are anchored to the fixed $T_b=80^\circ$C. From Equation \ref{eq:viscosity_law_note}, $\partial\ln\eta/\partial T|_{80^\circ\mathrm{C}} = -0.0503$ per $^{\circ}$C, giving $\Delta T_\nu = 19.9^\circ$C and $\Delta T_{\mathrm{eff}} \simeq 44.5^\circ$C. Equivalently, the sublayer comprises fluid with viscosity within a factor $e^{2.24}\simeq9.4$ of $\eta(T_b)$; because Equation \ref{eq:viscosity_law_note} is exponential of a quadratic function of $T$, rather than simple exponential, this 9.4 viscosity contrast criterion places the sublayer edge at $T^{*}=46^\circ$C ($\Theta=0.38$) rather than at $T_b-\Delta T_{\mathrm{eff}}=35.5^\circ$C ($\Theta=0.19$); the two constructions bracket the sublayer floor. In either case, neither $\Delta T_{\mathrm{eff}}$ nor the sublayer viscosity depends on the interior temperature, so this estimator predicts a strictly time-independent active-layer thickness for as long as the TBL edge remains colder than $T^{*}$ (interior $T_i$ below $31$--$42^\circ$C); once the interior warms past this point, the entire TBL lies within the mobile viscosity band and the estimator reduces to the endpoint average of Table~\ref{tab:relative_delta}.

This sublayer estimate is, moreover, the same calculation as the local-Rayleigh-number thickness used throughout this appendix. Marginal stability of the sublayer at $Ra_c=1708$, $\delta_{\mathrm{DJ}} =
[Ra_c\,\kappa\,\eta(T_b)/(\rho g \alpha\,\Delta T_{\mathrm{eff}})]^{1/3}$, differs from $\delta_{\mathrm{local}} = H(Ra_c/Ra_\delta)^{1/3}$ only in replacing the full temperature contrast $\Delta T = 55^\circ$C by $\Delta T_{\mathrm{eff}}$, both being anchored to the basal viscosity $\eta(T_b)$: $\delta_{\mathrm{DJ}} =
\delta_{\mathrm{local}}\,(\Delta T/\Delta T_{\mathrm{eff}})^{1/3} =
6.7~\mathrm{mm}\times(55/33.9)^{1/3} \simeq 7.8$~mm, or $7.1$~mm with $\Delta T_{\mathrm{eff}}=44.5^\circ$C. The two estimates bracket the observed $7.5$~mm, and the local-Rayleigh-number thickness of the main-text discussion section can be read as the \citeA{davaille1994onset} sublayer estimate evaluated with the full temperature contrast. The conclusion that the lower TBL does not thin as the interior warms therefore does not depend on the choice of effective-viscosity estimator.

The conductive heat flux across the lower TBL decreases because
\begin{equation}
    q_{\mathrm{TBL}}
    \sim
    k\frac{\Delta T_{\mathrm{TBL}}}{\delta}.
\end{equation}
Thus a decreasing bottom heat flux is fully compatible with an approximately constant TBL thickness. One cannot infer that the lower TBL becomes thinner from the decrease in $\Delta T_{\mathrm{TBL}}$ or heat flux alone. The viscosity evolution must also be included, and with the measured viscosity law of the present fluid, the expected thickness change is small; depending on the precise TBL definition, the TBL may even become slightly thicker at later times.

While the warmer fluid interior has little effect on the TBL thickness for a given $Ra_{\mathrm{TBL}}$, the effective critical $Ra_{\mathrm{TBL}}$ might be slightly lower after the first batch of plumes for two reasons. First, the overlying fluid becomes less viscous as its temperature increases. Over the temperature range in Table \ref{tab:relative_delta}, the viscosity contrast $\gamma$ between the overlying fluid and the TBL decreases from about 14 to 2 (so 7$\times$ relatively). According to the linear theory in \citeA{Stengel1982}, the effective critical Rayleigh number should change by less than 10\% over this range. If we use the effective viscosity contrast inferred in Section \ref{sec:spacing} based on minimum spacing over time and \citeA{whitehead1975dynamics}, i.e., 19-37 in stage 1 and 5 to 10 in stage 3, the relative change (4$\times$) is smaller. Even considering the maximum viscosity contrast in our experiment ($\gamma=65$ versus 1), the expected correction is still only about 15\%. \citeA{Stengel1982} formulated this result originally for a global Rayleigh number with an effective viscosity of the fluid, but it provides an order-of-magnitude estimate for the corresponding TBL correction. This small correction to the critical Rayleigh number has also been verified numerically by \citeA{Solomatov2006}.

The second possibility is the presence of finite-amplitude perturbations within the fluid introduced by the rising plumes \cite{white1988planforms}, compared with the very beginning of the experiment, when the fluid is nearly uniform and is being uniformly heated from below. However, such a change should not be significant, because the observed optical-distortion width of the plume stems remains around 5 mm throughout the experiment.

Together, these scaling arguments suggest that the TBL thickness should remain approximately constant as the fluid interior warms. The change in heat flux is mostly reflected in the decreasing size of the plume head, the width of each upwelling velocity region, and the rising velocity, rather than in the plume thickness.

\section{Comparison of plume head size with Whitehead and Luther (1975)}
\label{app:head}

\citeA{whitehead1975dynamics}'s Rayleigh-Taylor and source-fed head theory assumes that a low-viscosity spherical
cavity of radius $a$ grows from a localized source with volume flux $Q$ and detaches when
its Stokes rise speed exceeds its source-growth rate. We denote the dynamic viscosity of
the surrounding, more viscous upper fluid by $\eta_{\mathrm{upper}}$, and the density
contrast between the plume head and the surrounding fluid by $\Delta\rho$. In the
low-viscosity-head limit, the Stokes rise speed and the source-fed growth rate are
\begin{equation}
    v \simeq \frac{a^2 g\Delta \rho}{3\eta_{\mathrm{upper}}},
    \qquad
    \frac{da}{dt} = \frac{Q}{4\pi a^2}.
\end{equation}
Equating these two rates at detachment gives
\begin{equation}
    a_0 =
    \left(
        \frac{3Q\eta_{\mathrm{upper}}}{4\pi g\Delta\rho}
    \right)^{1/4}.
\end{equation}
Although our plumes are not injected from a prescribed source, an effective source flux can
be estimated from the volume of thermally anomalous boundary-layer fluid feeding a plume:
\begin{equation}
    Q_{\mathrm{eff}} \sim \frac{A_{\mathrm{eff}}\delta}{\tau}.
\end{equation}
Here $A_{\mathrm{eff}}$ is not necessarily the full plume-spacing cell area, because a newly
initiated head may drain only a local fraction of the thermal boundary layer. If
$A_{\mathrm{eff}}$ is approximated by an area of order the spacing-defined feeding cell,
$A_{\mathrm{eff}}\sim\lambda^2$, then for
$\lambda\simeq 0.08$--$0.12$ m, $\delta=6$--$7.5$ mm, and
$\tau=375$--$500$ s, and cold interior (25 $^\circ$C), the Whitehead and Luther spherical detachment diameter is of order $D_0=2a_0\simeq 14$--$21$ mm for the experimental viscosity and buoyancy. Late-stage heads can be even smaller as $\eta_{\mathrm{upper}}$ decreases. Here $\Delta\rho$ is based on $\Delta T=30\sim40 ^\circ$C from a half space cooling temperature profile after heating for 375 to 500 s, cf. \ref{app:TBL}, but the result will not change much using the full experimental $\Delta T=55^\circ$C. Using the determined stage-1 spacing and $\Delta T=55^\circ$C, $D_0=2a_0\simeq 16.4$--$18.7$ mm. This range includes the newly identified just-initiated plume head diameter of $\sim18$ mm, suggesting that the Whitehead and Luther spherical head scale is appropriate for the initial head just above the basal thermal boundary layer. This comparison should not be interpreted as a prediction of the latter plume head diameter at mid-tank or near the upper boundary, because those heads have continued to rise, feed, deform, entrain ambient fluid, and thermally diffuse after detachment. It is also important to note that the Whitehead and Luther analytical head radius is a spherical-equivalent scale; a mushroom-shaped, flattened, or laterally spreading head with the same volume can have a larger apparent horizontal diameter.

A separate first-order estimate of the later, mature head-size scale can be obtained from a stage-averaged feeding-volume argument. If a plume drains a spacing-defined area of order $\lambda^2$ from a thermal boundary layer of thickness $\delta$, the available thermal volume is
\begin{equation}
    V_{\mathrm{feed}} \sim f\lambda^2\delta,
\end{equation}
where $f\leq1$ accounts for incomplete collection into the coherent head, thermal diffusion, and possible distinction between thermal and material boundaries. The corresponding spherical-equivalent diameter is
\begin{equation}
    D_{\mathrm{scale}} \sim
    \left(\frac{6f\lambda^2\delta}{\pi}\right)^{1/3}.
\end{equation}
For $f=1$ and $\delta=6$--$7.5$ mm, and spacing $\lambda\simeq 0.08$--$0.12$ m, this gives head-size scales of order $42$--$60$ mm across the observed plume-spacing range, comparable to the observed mid-tank head sizes (60 mm for the first batch and 40 mm for the middle and end of the experiment). We view this only as a first-order, stage-averaged estimate, because $\lambda$ is itself a stage-averaged spacing and the true drainage area of any individual plume evolves through time and is modified by plume--plume interaction. Nevertheless, as we obtained the spacing of plumes at varies heights, effectively the mean spacing reflects a mid-tank height result, therefore comparing the inferred head size with the observed mid-tank height plume head size is appropriate. 

The resulting head/stem ratio depends on which head stage is considered. The newly
initiated head, $D_0\simeq18$ mm, gives
\begin{equation}
    \frac{D_0}{\delta} \simeq 3.0,\ 2.6,\ \mathrm{and}\ 2.4
\end{equation}
for $\delta=6,\ 7,\ \mathrm{and}\ 7.5$ mm, respectively. Later mid-tank heads are larger. Using the later-stage typical value $D_{\mathrm{head}}\sim40$ mm gives
\begin{equation}
    \frac{D_{\mathrm{head}}}{\delta}
    \simeq 6.7,\ 5.7,\ \mathrm{and}\ 5.3
\end{equation}
for $\delta=6,\ 7,\ \mathrm{and}\ 7.5$ mm. Using the earlier mid-tank value $D_{\mathrm{head}}\sim60$ mm gives ratios of
\begin{equation}
    \frac{D_{\mathrm{head}}}{\delta}
    \simeq 10.0,\ 8.6,\ \mathrm{and}\ 8.0.
\end{equation}
The just-initiated head/stem ratio is close to the classical source-fed initial-head scale, while the mature mid-tank heads are substantially larger because they have continued to feed, deform, and spread after initiation. These mature ratios are larger than, but still broadly comparable to, the head/stem ratios of $\sim3$--$5$ reported by
\citeA{lithgow2001plume} for thermal plumes over $Ra\sim10^5$--$10^8$, where the
ratio increases with increasing $Ra$. The larger ratios in our experiment are therefore
consistent with the high effective Rayleigh number of the hot basal boundary layer and with continued head growth after detachment from the basal TBL.

%
%

\bibliography{agusample.bib}

%
%
%
%
%

\end{document}